\documentclass[twocolumn,twocolappendix]{aastex631}
\graphicspath{{./fig/}{./png/}}

\newcommand{\EQ}{\begin{equation}}
\newcommand{\EN}{\end{equation}}
\newcommand{\EQA}{\begin{eqnarray}}
\newcommand{\ENA}{\end{eqnarray}}

\newcommand{\Eq}[1]{Equation~(\ref{#1})}
\newcommand{\Eqs}[2]{Equations~(\ref{#1}) and~(\ref{#2})}

\newcommand{\Sec}[1]{Section~\ref{#1}}

\newcommand{\Fig}[1]{Figure~\ref{#1}}

\newcommand{\Figp}[2]{Figure~\ref{#1}({#2})}

\newcommand{\Tab}[1]{Table~\ref{#1}}

\newcommand{\App}[1]{Appendix~\ref{#1}}

\newcommand{\bra}[1]{\langle #1\rangle}

{}
{}

\newcommand{\tildeAA}{\tilde{\mathbf{A}}}
\DeclareMathAlphabet\mathbfcal{OMS}{cmsy}{b}{n}
\newcommand{\calAA}{{\mathbfcal{A}}}
\newcommand{\tildecalAA}{\tilde{\mathbfcal{A}}}
\newcommand{\tildeBB}{\tilde{\mathbf{B}}}
\newcommand{\tildeEE}{\tilde{\mathbf{E}}}

\newcommand{\kk}{\mathbf{k}}

\newcommand{\xx}{\mathbf{x}}

\newcommand{\BB}{\mathbf{B}}

\newcommand{\EE}{\mathbf{E}}
\newcommand{\JJ}{\mathbf{J}}

\newcommand{\AAA}{\mathbf{A}}

\newcommand{\uu}{\mathbf{u}}

\newcommand{\FFF}{\mbox{\boldmath ${\cal F}$} {}}

\newcommand{\nab}{{\mathbf{\nabla}}}

\newcommand{\MMMM}{\mbox{\boldmath ${\sf M}$} {}}

\newcommand{\SSSS}{\mbox{\boldmath ${\sf S}$} {}}

\newcommand{\eee}{{\sf e}}

\newcommand{\TTT}{{\sf T}}
\newcommand{\tildeh}{\tilde{h}}
\newcommand{\tildeT}{\tilde{T}}

\newcommand{\tildeTTT}{\tilde{\sf T}}
\newcommand{\ii}{{\rm i}}

\newcommand{\dd}{{\rm d} {}}

\def\H{\mbox{\rm H}}

\def\Sp{\mbox{\rm Sp}}

\def\EEM{{\cal E}_{\rm M}}
\def\EEEM{{\cal E}_{\rm EM}}
\def\EEEl{{\cal E}_{\rm E}}

\def\EEGW{{\cal E}_{\rm GW}}
\def\OmGW{{\Omega}_{\rm GW}}

\def\EGW{E_{\rm GW}}

\def\EK{E_{\rm K}}
\def\EM{E_{\rm M}}

\def\hrms{h_{\rm rms}}

\def\kpeak{k_{\rm *}}

\def\EM{E_{\rm M}}

\def\Brms{B_{\rm rms}}

\def\half{{\textstyle{1\over2}}}

\def\onethird{{\textstyle{1\over3}}}

\newcommand{\W}{\,{\rm W}}
\newcommand{\V}{\,{\rm V}}

\newcommand{\T}{\,{\rm T}}
\newcommand{\G}{\,{\rm G}}

\newcommand{\uHz}{\,\mu{\rm Hz}}
\newcommand{\nHz}{\,{\rm nHz}}

\newcommand{\K}{\,{\rm K}}
\newcommand{\g}{\,{\rm g}}
\newcommand{\s}{\,{\rm s}}

\newcommand{\GeV}{\,{\rm GeV}}
\newcommand{\MeV}{\,{\rm MeV}}

\newcommand{\m}{\,{\rm m}}

\newcommand{\J}{\,{\rm J}}

\newcommand{\A}{\,{\rm A}}

\begin{document}

\title{Simulating relic gravitational waves from inflationary magnetogenesis}

\correspondingauthor{Axel Brandenburg}
\email{brandenb@nordita.org}
\author[0000-0002-7304-021X]{Axel Brandenburg}
\affiliation{Nordita, KTH Royal Institute of Technology and Stockholm University,
Hannes Alfv\'ens v\"ag 12, SE-10691 Stockholm, Sweden}
\affiliation{Department of Astronomy, AlbaNova University Center,
Stockholm University, SE-10691 Stockholm, Sweden}
\affiliation{McWilliams Center for Cosmology \& Department of Physics,
Carnegie Mellon University, Pittsburgh, PA 15213, USA}
\affiliation{School of Natural Sciences and Medicine, Ilia State University,
3-5 Cholokashvili Avenue, 0194 Tbilisi, Georgia}

\author[0000-0002-2549-6861]{Ramkishor Sharma}
\affiliation{Inter University Centre for Astronomy and Astrophysics,
Post Bag 4, Pune University Campus, Ganeshkhind, Pune 411 007, India}

\begin{abstract}
We present three-dimensional direct numerical simulations of the production of magnetic
fields and gravitational waves (GWs) in the early Universe during a low energy scale
matter-dominated post-inflationary reheating era, and during the early
subsequent radiative era, which is strongly turbulent.
The parameters of the model are determined such that it avoids a number of
known physical problems and produces magnetic energy densities between
0.03\% and 0.5\% of the critical energy density at the end of reheating.
During the subsequent development of a turbulent magnetohydrodynamic
cascade, magnetic fields and GWs develop a spectrum that extends to
higher frequencies in the millihertz (nanohertz) range for models with
reheating temperatures of around $100\GeV$ ($150\MeV$) at the beginning
of the radiation-dominated era.
However, even though the turbulent cascade is fully developed, the GW
spectrum shows a sharp drop for frequencies above the peak value.
This suggests that the turbulence is less efficient in driving GWs
than previously thought.
The peaks of the resulting GW spectra may well be in the range accessible
to space interferometers, pulsar timing arrays, and other facilities.
\end{abstract}


\section{Introduction}

During the past few years, numerical simulations of gravitational wave
(GW) generation from early Universe turbulence have become an essential tool
in predicting the stochastic background that the Laser Interferometer
Space Antenna \citep[LISA; see, e.g.,][]{2017arXiv170200786A} and other space
interferometers \citep[e.g.,][]{2021CmPhy...4...34T} might see in the future.
Most of the existing predictions are based on analytical models
\citep{2002PhRvD..66j3505D, kosowsky2002, caprini2006,sigl2018},
which tend to make
simplifying assumptions about the nature of turbulent wave generation;
see \cite{2016JCAP...04..001C} and \cite{2018CQGra..35p3001C} for recent
reviews emphasizing the feasibility and prospects of observing such relic GWs.

A particularly popular source of turbulence in the early Universe is
the electroweak phase transition.
\cite{2015PhRvD..92l3009H,2017PhRvD..96j3520H} have produced numerical
simulations of GW generation by assuming a first order phase transition
\citep{kosowsky1992,1994PhRvD..49.2837K,nicolis2004,ellis2019,ellis2020}.
Even if the phase transition is not a first order one, as initially
assumed, it is still possible to produce primordial turbulence from
magnetic fields that could be generated during various epochs in the early
Universe \citep[see, e.g.,][]{cornwall1997,joyce1997,arun2016,2018PhRvL.121b1301M}.
The existence of large-scale magnetic fields in the early universe
is motivated by indirect evidence of their presence in the intergalactic regime
from the non-detection of GeV photons in blazar observations
\citep{neronov,taylor2011,tavecchio2011,ackermann2018,Archambault_2017}.

Both analytical considerations 
\citep{2002PhRvD..66j3505D,kosowsky2002,2007PhRvD..76h3002G,tina2008}
and numerical simulations \citep{2020PhRvD.102h3512R}
have demonstrated that there can be a direct correspondence between the
turbulence spectrum and the resulting GW spectrum.
An important additional property of GWs might be their circular
polarization, which could be caused by 
helical turbulence \citep{2005PhRvL..95o1301K} or by helical magnetic fields
\citep{namba2016,sigl2018,anand2018,2020PhRvD.101j3526S,2021JCAP...03..026O}.
Again, numerical simulations have confirmed the direct correspondence
between the fractional helicity of magnetic fields and the resulting
circular polarization of GWs \citep{2021PhRvR...3a3193K}.

Magnetogenesis during quantum chromodynamic (QCD) phase
transitions \citep{1989ApJ...344L..49Q,sigl1997,tevzadze2012}
provide another possible
avenue for GW generation at low frequencies in the nanohertz range
\citep{tina2010,2021PhRvD.103L1302N}.
If the characteristic scale of QCD turbulence is a significant fraction
of the Hubble horizon at that time, as suggested by some models
\citep[e.g.,][]{2005PhRvD..71f5017K}, the resulting GW spectrum could show
a marked drop in the spectral energy density for frequencies above the value typical
of the turbulent driving scale \citep{2021arXiv210212428B}.
This result has been obtained by assuming the turbulence to be
driven by a monochromatic forcing function.
However, it remains unclear how sensitive such results are to the
assumption of an artificially adopted forcing function.
For this reason, it is essential to include the magnetogenesis mechanism
in the actual simulations of GW production, without using any artificial forcing.
One such mechanism is the dynamo effect associated with the chirality
of fermions \citep{joyce1997}, which is referred to as
the chiral magnetic effect \citep{1980PhRvD..22.3080V}.
Numerical simulations of the resulting GW generation predict that
their power depends on the speeds of magnetic field generation
and saturation \citep{2021ApJ...911..110B}.
However, this model suffers from the difficulty that the typical
length scale associated with the chiral magnetic effect is very short.
For this reason, we focus here on magnetogenesis during inflation.
It is traditionally expected to produce a large-scale magnetic field
\citep{1992ApJ...391L...1R,martin2008,kandu2010,2017JCAP...12..002K,fujita2019}.
However, as we discuss next, it is unclear how to properly model GW
production for such a magnetic field.

Earlier approaches modeled the process of GW production from magnetic
fields by assuming a magnetic field to be given; see the
models ini1--ini3 in \cite{2020PhRvD.102h3512R} for such examples.
However, this corresponds to switching on a magnetic field abruptly at
a particular time.
Therefore, the process of switching on a magnetic field with a given
spectrum played a decisive role in the resulting GW energy and strain.
The result would be different if the field was gradually being produced
by some magnetogenesis mechanism.
Including a suitable magnetogenesis model is what will be presented
in this paper.
This will then also allow us to quantify the resulting differences.
We can therefore demonstrate what difference it would make when we just
switched on the magnetic field from our magnetogenesis simulation without
including the corresponding GW generation until that time.

A popular model of inflationary magnetogenesis is that of
\cite{1992ApJ...391L...1R}, where electromagnetic fields originate
from quantum fluctuations \citep{1985NuPhB.259..730F} that are being
amplified during inflation owing to the breaking of conformal invariance
\citep{1988PhRvD..37.2743T,1993PhRvD..48.2499D}.
This is achieved by a suitable coupling of the inflaton field to the
electromagnetic field through a function $f$ leading to a term of the form
$f^2 F^{\mu\nu} F_{\mu\nu}$ in the Lagrangian density, where $F_{\mu\nu}$
is the Faraday tensor.
One usually assumes $f$ to be proportional to some power $\alpha$
of the scale factor $a$ \citep{bamba2007,martin2008,kandu2010}, i.e.,
\EQ
f(a)\propto a^\alpha.
\label{fa_alpha}
\EN
Of particular interest are scale-invariant magnetic fields, which
can be obtained for $\alpha=2$ and $-3$ assuming constant expansion rate during inflation.
Although the strength of the resulting magnetic field in such models
may well be of astrophysical interest,
they suffered from three major shortcomings.
(i) In the case $\alpha=2$, the function $f$ increases from a certain initial
value to a very large value at the end of inflation.
Demanding standard electromagnetism at the end of inflation requires
$f=1$, but this would imply a very small value of $f$ 
at the beginning of inflation.
This results in very large values of the effective electric charge,
defined as $e_{\rm eff}=e/f^2$ \citep{kandu2010,Kobayashi14}, where $e$ is the
standard elementary charge.
Therefore, there will be a very large coupling between the
electromagnetic and charged fields at the beginning of inflation.
For this reason, this theory would be in the nonperturbative regime
and would therefore not be reliable.
This is known as the strong coupling problem \citep{mukhanov2009}.
(ii) In the other case, where $\alpha=-3$, the electric energy density
diverges and may overshoot the background energy density during inflation.
This problem is known as the backreaction problem \citep{mukhanov2009}.
(iii) The production of charged particles in the presence of strong
electric fields due to the Schwinger effect can lead to a premature
increase in the electric conductivity, which shorts the electric field and
prevents further magnetic field growth; see \cite{2014JHEP...10..166K}.
This problem also applies to models that solve the backreaction problem
by choosing a low energy scale inflation \citep{2013JCAP...10..004F}.
Such a problem could be avoided if charged particles get sufficiently
large masses by some mechanism in the early Universe, as suggested by
\cite{Kobayashi+Sloth19}.

The \cite{2017PhRvD..96h3511S,2018PhRvD..97h3503S} model
addresses the three problems\footnote{
\cite{2017PhRvD..96h3511S,2018PhRvD..97h3503S} discuss the Schwinger
effect constraint during inflation, but not in the post-inflationary
matter-dominated era.
The reason is that a calculation of the conductivity due to the Schwinger
mechanism in a matter-dominated Universe has yet to be done.
If one considers instead the expression for the conductivity
in a de Sitter spacetime, given by \cite{2014JHEP...10..166K} and
\cite{Kobayashi+Sloth19}, it appears that the Schwinger effect constraint
becomes important in the last phase of the matter-dominated era.
However, more meaningful conclusions require a detailed investigation.}
by constraining the form of $f(a)$ such that $\alpha=2$ during inflation,
starting at an initial value of unity, thereby solving the strong coupling
problem, and to obey
\EQ
f(a)\propto a^{-\beta}
\label{fa_beta}
\EN
during a post-inflationary era, which is assumed to be matter dominated.
The exponent $\beta>0$ is calculated for a given reheating temperature.
By choosing the reheating temperature to be at the electroweak scale
of $100\GeV$ and the total electromagnetic energy density to be $1\%$
of the background energy density, the number of $e$-folds of the scale
factor during inflation, $N$, and during reheating, $N_{\rm r}$, are
found to be 34 and 9.3, respectively.
To arrive at standard electrodynamics with $f=1$ at the end of
reheating, we require $\alpha N=\beta N_{\rm r}$, and therefore
$\beta=\alpha N/N_{\rm r}=7.3$.
Alternatively, following \cite{2020PhRvD.101j3526S} and
\cite{2021arXiv210209358S}, we can consider
the case where the reheating temperature is at the QCD energy scale
of $150\MeV$, for which one finds $N=36$, $N_{\rm r}=28$, and therefore
$\beta=2.7$; see \App{ModelDetails} for the details.

A similar model for helical magnetic field generation and polarized GWs
was recently considered by \cite{2020PhRvD.101j3526S} and \cite{2021JCAP...03..026O}.
These studies are based on earlier work of \cite{rajeev2010},
\cite{2014JCAP...10..056C}, \cite{fujita2015},
\cite{2018PhRvD..97h3503S}, and \cite{fujita2019}.
However, the numerical consideration of such models is beyond the
scope of the present work and is the subject of a separate paper
\citep{Bran+He+Shar21}.

Here, we adopt the aforementioned magnetogenesis model of
\cite{2017PhRvD..96h3511S} to compute both
electromagnetic fields and GWs resulting from the electromagnetic
stress during the late reheating phase, when the conductivity is still
negligible, and during the early radiation-dominated phase when the
conductivity is high and the laws of magnetohydrodynamics (MHD) are
applicable.
In the first step, magnetic fields and GWs exist only on very large
length scales.
The significance of the second step is therefore to produce magnetic
fields and GWs at smaller length scales through turbulence that is being
driven by the Lorentz force from electric currents once the conductivity
is high.
By making simplifying assumptions, \cite{2020PhRvD.101j3526S} have
already considered this case, but avoiding the restrictions resulting
from these assumptions requires self-consistently computed turbulence,
which can only be done numerically.

\section{The model}
\label{TheModel}

We consider a periodic domain of size $L^3$.
The smallest wavenumber is then $2\pi/L\equiv k_1$.
In this work, we use cubic domains with $n=512$ or $1024$ mesh points
in each direction with $k_1=1$, so the Nyquist wavenumber,
$k_{\rm Ny}=k_1 n/2$, is either $256$ or $512$.
We adopt a spatially flat Friedmann--Lema\^itre--Robertson--Walker metric.
Throughout this paper, we use conformal time $\eta=\int \dd t/a(t)$,
where $t$ is the physical time, and work with comoving variables that
are scaled by the appropriate powers of $a$.
In particular, the MHD equations then become equal to the MHD
equations in a non-expanding universe \citep{1996PhRvD..54.1291B}.
Our comoving variables therefore describe the departures from the
expansion of the Universe.
The speed of light is always set to unity and the Lorentz--Heaviside
unit system is used for the Maxwell equations.
We also set the density at the beginning of the radiation-dominated
era to unity.
This also implies that the mean radiation energy density is unity and
therefore the magnetic energy densities quoted below are automatically
the fractional magnetic energy densities with respect to the 
radiative energy density.

As was suggested in the introduction, we perform the simulations
in two separate steps.
In step~I, during the end of reheating, we solve the Maxwell equations
with zero conductivity, but, owing to the breaking of conformal
invariance, with a nonvanishing $f''/f$ term, where primes denote
derivatives with respect to $\eta$.
Similarly, in the GW equation, the $a''/a$ term is nonvanishing.
In step~II, we assume a rapid transition into the radiation-dominated
era, where the electric field can be neglected and we thus
solve the MHD equations.
Owing to finite conductivity, electric currents can flow
and drive fluid motions through the Lorentz force,
which leads to additional induction and magnetic
field amplification at small length scales.
In that case, $f=1$ and $a\propto\eta$ grows linearly,
so $f''/f=a''/a=0$.

Following \cite{2020PhRvD.102h3512R}, we scale the conformal time
at the beginning of the radiative era to unity, i.e., $\eta=1$.
We simulate the last phase $\eta_{\rm ini}\leq\eta\leq1$
of the reheating interval, where the scale factor is taken
to be $a=(\eta+1)^2/4$ so as to match $a=1$ at $\eta=1$
\citep{2017PhRvD..96h3511S}.
In practice, we consider as the initial value of the conformal time
$\eta_{\rm ini}=-0.9$, corresponding to an initial scale factor of
$a_{\rm ini}=1/400$.
Thus, we have
\EQ
\frac{a''}{a}=\frac{2}{(\eta+1)^2}\quad\mbox{and}\quad
\frac{f''}{f}=\frac{2\beta(2\beta+1)}{(\eta+1)^2}.
\label{d2f}
\EN
In step~I, we solve the following equations for variables in Fourier
space, denoted by tildae on the scaled magnetic vector potential $\calAA$
and the strains $h_+$ and $h_\times$ for the two linear polarizations modes:
\EQ
\tildecalAA''+\left(\kk^2-\frac{f''}{f}\right)\tildecalAA=0,
\label{dAk2dt2}
\EN
\EQ
\tildeh_{+/\times}''
+\left(\kk^2-\frac{a''}{a}\right)\tildeh_{+/\times}
={6\over a}\,\tildeT_{+/\times}.
\label{d2hdt2}
\EN
Here, $\tildecalAA=f\tildeAA$, where $\tildeAA$ is the
magnetic vector potential and $\tildeT_{+/\times}(\eta,\kk)=
\eee_{+/\times}^{ij}\tildeTTT_{ij}(\eta,\kk)$ are the
$+$ and $\times$ polarizations of the traceless-transverse
projected stress in Fourier space, where
$\tildeTTT_{ij}(\eta,\kk)=\int {\TTT}_{ij}(\eta,\xx)\, e^{-\ii\kk\cdot\xx}\dd^3\xx$
is the Fourier transformation of the electromagnetic stress,
given in real space by
\EQ
{\TTT}_{ij}=f^2\,(B_i B_j+E_i E_j).
\label{TEB}
\EN
Here, $\EE=-\partial\AAA/\partial\eta$ and $\BB=\nab\times\AAA$ are
computed through inverse Fourier transformation, $\EE(\eta,\xx)=\int
\tildeEE(\eta,\kk)\, e^{\ii\kk\cdot\xx}\dd^3\kk/(2\pi)^3$,
and likewise for $\tildeBB(\eta,\kk)$, which is given by
$\tildeBB=\ii\kk\times\tildeAA$.

As initial condition for $\eta=\eta_{\rm ini}$, we employ a random,
Gaussian-distributed magnetic field with a magnetic energy spectrum
$\EM(k)\propto k^3$ (for $k<\kpeak$) and $\propto k^{1-4\beta}$
(for $k>\kpeak$), where
\EQ
\kpeak(\eta)=\sqrt{2\beta(2\beta+1)}/(\eta+1)
\label{kpeak}
\EN
is evaluated at $\eta=\eta_{\rm ini}$; see \App{InitCond}.
This implies that the magnetic and electric energy spectra peak
at a wavenumber that lies well within the computational domain, i.e.,
$k_1<\kpeak(\eta)<k_{\rm Ny}$.
The magnetic energy spectrum is normalized such that
$\bra{\BB^2}/2\equiv\EEM(\eta)=\int\EM(\eta,k)\,\dd k$.
It is important to emphasize that $\kpeak$ is sufficiently
far away from the minimal and maximal wavenumbers available in our
simulation, so the $k$-integrated spectral energy densities are
not sensitive to our precise choice of domain size and resolution.
We denote the spectra of the quantity $\BB$ through integration over
concentric shells in wavenumber space as $\Sp(\BB)/2\equiv\EM(k)$.
Likewise, the electric spectrum is $\Sp(\EE)/2\equiv E_{\rm E}(k)$
with $\bra{\EE^2}/2\equiv\EEEl=\int E_{\rm E}(k)\,\dd k$.
The GW energy spectrum is $\EGW(k)=[\Sp(h_+)+\Sp(h_\times)]/6$
in our normalization, where the critical energy density is unity;
see also \cite{2020PhRvD.102h3512R}.
In step~II, we also present kinetic energy spectra, which are
defined as $\EK(k)=\Sp(\uu)/2$.
Fluctuations of the radiation energy density are ignored.
We recall in this connection that the mean radiation
energy density is normalized to unity.

Owing to the rapid increase of the spectra at small $k$, the detailed
initialization of $\tildeEE$ turns out not to be critical and it
suffices to initialize the electric field such that it is a solution
to the electromagnetic wave equation with $\EE=-\AAA'$,
and therefore $\tildeEE=\ii k\tildeAA$,
where $k=|\kk|$ is the length of the wavevector.
This implies that $E_{\rm E}(k)$ is initially equal to the magnetic energy
spectrum at all $k$, i.e., $\EM(k)=E_{\rm E}(k)$ for $\eta=\eta_{\rm ini}$.
These spectra then begin to change and grow rapidly at small
wavenumbers, but $\EM(k)$ retains its initial $k^3$ scaling and
$E_{\rm E}(k)$ attains a $k^1$ scaling.
This is because the $f''/f$ term in \Eq{dAk2dt2} now dominates
over the $\kk^2$ term at small $k$, so there is no longer the
$k$-dependent factor between $\tildeAA$ and $\tildeEE$.
The electric field spectrum then becomes proportional
to the spectrum of the vector potential, and therefore
\EQ
E_{\rm E}(k,\eta)\propto k^{-2}\EM(k,\eta)\quad\mbox{for $k<\kpeak(\eta)$}.
\EN
Since \Eq{dAk2dt2} is linear, we can easily find by
trial and error the magnetic energy that is needed so
that at $\eta=1$, the mean electromagnetic energy density,
\EQ
\EEEM\equiv\EEEl+\EEM=\bra{\EE^2+\BB^2}/2,
\EN
is a few the percent of the radiation energy density.

\begin{figure*}\begin{center}
\includegraphics[width=\textwidth]{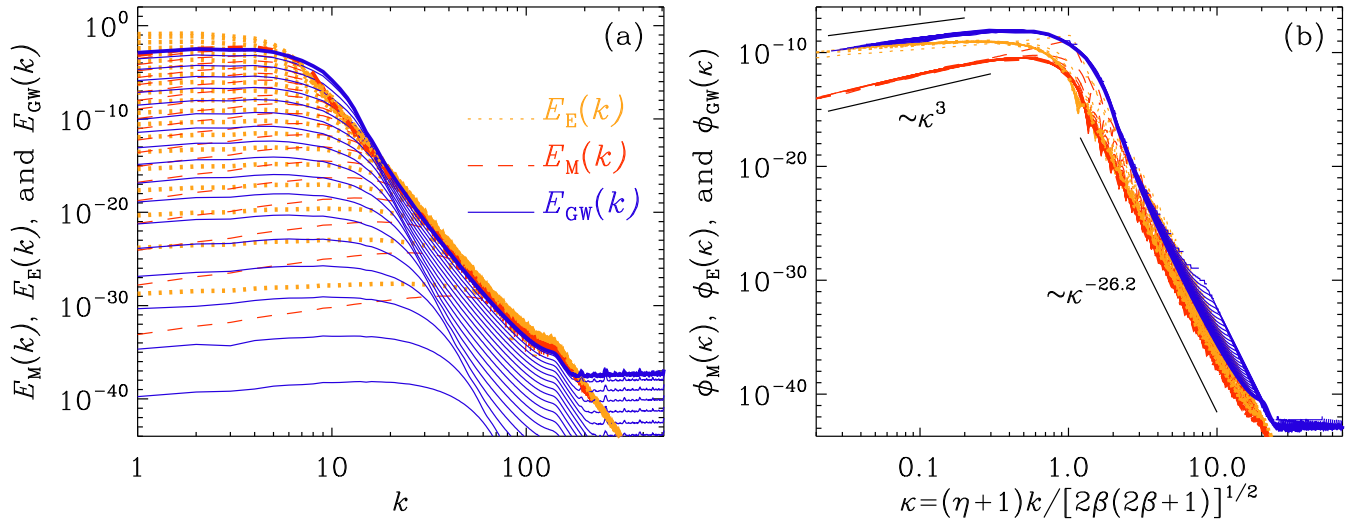}
\end{center}\caption{
(a) $\EM(k)$ (red lines), $E_{\rm E}(k)$ (orange lines), and
$\EGW(k)$ (blue lines) for Run~A1.
(b) $\phi_{\rm M}(\kappa)$ (red lines), $\phi_{\rm E}(\kappa)$ (orange lines), and
$\phi_{\rm GW}(\kappa)$ (blue lines).
The black straight lines denote the slopes
$\EM(k)\propto \kappa^3$ and $\propto \kappa^{1.7}$ for small $\kappa$.
}\label{rspec34_1024b68d}\end{figure*}

In step~II, for $\eta>1$, the conductivity $\sigma$ is finite and so the
evolution of $\EE$ can be omitted and the magnetic and GW fields are
evolved by solving the MHD and GW equations, as described in previous
papers \citep{2020GApFD.114..130R,2020PhRvD.102h3512R}, where the
evolution equation for $\AAA$,
\EQ
\frac{\partial\AAA}{\partial\eta}=
\uu\times\BB+\sigma^{-1}\nabla^2\AAA
\label{dAdt}
\EN
is solved in real space.
This equation includes the induction effects from the
velocity and the finite conductivity.
It is solved together with \citep{1996PhRvD..54.1291B}
\begin{equation}
\frac{\partial\uu}{\partial\eta}=
-\uu\cdot\nab\uu-{1\over4}\nab\ln\rho
+{3\over4\rho}\JJ\times\BB
+\FFF_\nu+\FFF,
\label{dudt}
\end{equation}
\begin{equation}
\frac{\partial\ln\rho}{\partial\eta}=
-\frac{4}{3}\left(\nab\cdot\uu+\uu\cdot\nab\ln\rho\right) + {\cal H},
\label{dlnrhodt}
\end{equation}
where $\FFF=(\nab\cdot\uu+\uu\cdot\nab\ln\rho)\uu/3 
-[\uu\cdot(\JJ\times\BB)+\JJ^2/\sigma]\uu/\rho$,
and ${\cal H}=[\uu\cdot(\JJ\times\BB)+\JJ^2/\sigma]\rho$
are additional terms that are retained in the calculation,
and $\FFF_\nu=2\nab\cdot(\rho\nu\SSSS)/\rho$ is the viscous force,
where ${\sf S}_{ij}=\half(u_{i,j}+u_{j,i})-\onethird\delta_{ij}\nab\cdot\uu$
are the components of the rate-of-strain tensor with commas denoting partial
derivatives, and $\nu$ is the kinematic viscosity.
In all cases considered below, we assume a magnetic Prandtl number
of unity, i.e., $\nu\sigma=1$.
We recall that $a=\eta$ during the radiation-dominated phase,
and therefore we have $a''/a=0$ in \Eq{d2hdt2}.

In step~II, the stress associated with the electric field is absent
in the expression for $\TTT_{ij}$.
Instead, the Reynolds stress $\gamma^2\rho u_i u_j$ now enters.
Here, $\gamma$ is the Lorentz factor with $\gamma^2=1/(1-\uu^2)$.

For both steps~I and II, we use the {\sc Pencil Code}
\citep{2021JOSS....6.2807P}, which is primarily designed to solving
large sets of partial differential equations on massively parallel
computers using sixth-order finite differences and a third-order time
stepping scheme.
However, the code is versatile and allows the GW equations to be advanced
analytically in Fourier space from one time step to the next; see the
detailed description in \cite{2020GApFD.114..130R}.
In step~I, we solve \Eq{dAk2dt2} in a similar fashion as \Eq{d2hdt2},
where the time advance from one time step to the next is done
analytically; see \App{MaxwellEquation}, where we describe the more
general case with $\sigma\neq0$.
It should be emphasized that, although \Eqs{dAk2dt2}{d2hdt2}
are linear for $\tildeAA$ and $\tildeh_{+/\times}$, respectively,
the combined problem is not because $\tildeT_{+/\times}(\eta,\kk)$
depends quadratically on $\EE$ and $\BB$ through \Eq{TEB}.

In some of our simulations, the electromagnetic energy density exceeds
10\% of the radiation energy density.
This would be unrealistically large and those cases are only included
for comparison with others of smaller electromagnetic energy density.
Indeed, it would then no longer be obvious that the linearized GW
equations are still applicable and that quadratic terms can be neglected.
Although the fractional GW energy densities are always much below the
fractional electromagnetic energy densities, it is conceivable that
nonlinear effects could play a role in certain wavenumber ranges.
The {\sc Pencil Code} does allow for such nonlinear effects in the GW
field to be incorporated.
Preliminary studies suggest that nonlinear contributions to the stress
begin to enhance the resulting GW energy spectra at large wavenumbers
when the electromagnetic energy density reaches about 30\%
of the radiation energy density.
However, such cases are not included in the present study and their
details will be presented elsewhere.

\section{Results}

\subsection{Evolution during step~I}
\label{EvolStepI}

During reheating, the energies of various fields increase rapidly
in power-law fashion, i.e., ${\cal E}_i(\eta)\propto(\eta+1)^{p_i}$
with $i={\rm M}$, ${\rm E}$, or ${\rm GW}$ for the magnetic, electric,
and GW energies, respectively.
Analytically, as shown in \App{InitCond}, we expect
$p_{\rm M}=p_{\rm E}=4\beta-2$, which is $27.2$, $25.2$, and $8.8$
for $\beta=7.3$, $6.8$, and $2.7$, respectively.
The growth continues to occur at progressively smaller wavenumbers.
The results are qualitatively similar for $\beta=2.7$, which is relevant
to reheating at the QCD energy scale; see \App{ModelDetails}.
At all larger wavenumbers, the magnetic field oscillates in space
and time, but does not increase on average.
The GW field evolves in a similar fashion, but even more rapidly,
and empirically with $p_{\rm GW}=2(p_{\rm M}-1)$.

\begin{figure*}\begin{center}
\includegraphics[width=\textwidth]{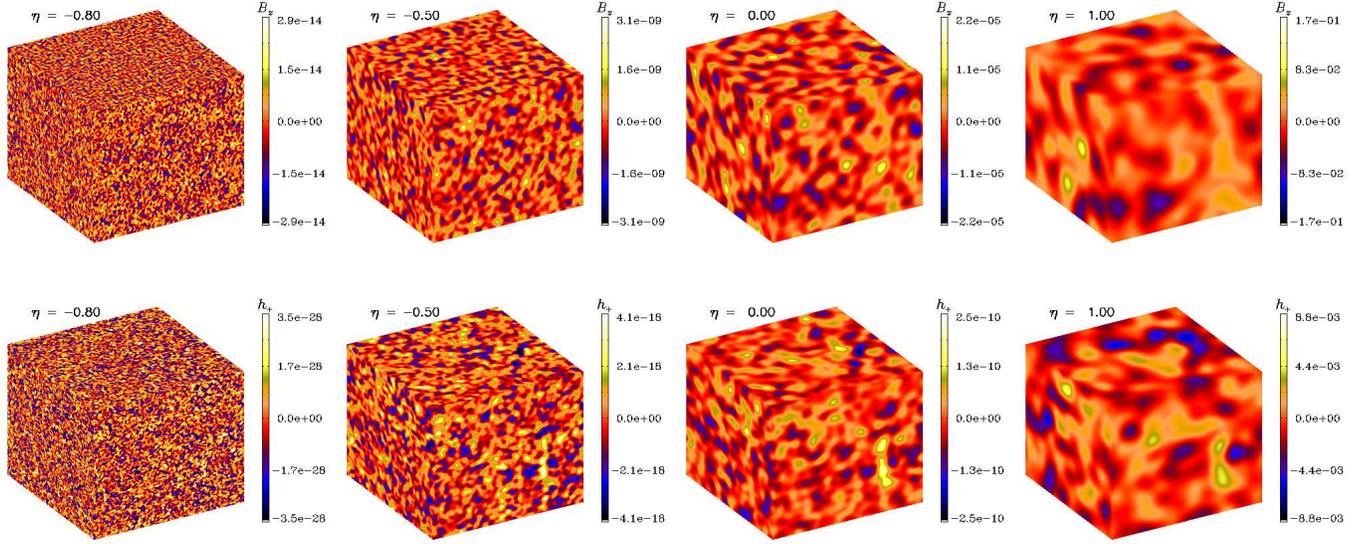}
\end{center}\caption{
Visualizations of $B_z$ (top) and $h_+$ (bottom) on the periphery of
the computational domain for Run~A1 at $\eta=-0.8$, $-0.5$, $0$, and $1$
during step~I.
The color scale is symmetric about zero and adjusted with respect to
the instantaneous extrema.
}\label{AB_1024b68d}\end{figure*}

In \Figp{rspec34_1024b68d}{a}, we show for the case with
$\beta=6.8$ magnetic, electric, and GW energy spectra in regular time intervals.
The spectra collapse on top of each other when plotting them versus
\EQ
\kappa(\eta)=(\eta+1) k/[2\beta(2\beta+1)]^{1/2} \equiv k/\kpeak(\eta)
\EN
and multiplying by a compensating factor $(\eta+1)^{-(p_i+1)}$.
Thus, we define (for $\eta\leq1$)
\EQ
\phi_i(\kappa)=(\eta+1)^{-(p_i+1)}E_i(k,\eta).
\EN
This implies $\EEM(\eta)\equiv\int E_M(k,\eta)\,\dd k
\propto(\eta+1)^{p_{\rm M}}$ for the temporal growth of the
($k$-integrated) magnetic energy density for $\eta<1$.
In \Fig{AB_1024b68d}, we show visualizations of $B_z$ and $h_+$ on the
periphery of the computational domain for Run~A1.
We see that the typical length scales of both fields increase with time.
This is due to the fact that the destabilizing term $f''/f$ in
\Eq{dAk2dt2} decreases with time and remains important only on
progressively larger length scales; see \Eq{d2f}.

We have also inspected visualizations of $\dot{h}_+$ and found that they
looked virtually identical to those of $h_+$.
Unlike step~II, where this is not the case (discussed below), we have
therefore not shown $\dot{h}_+$ here.
However, we have looked at the local correlation between the two
for each mesh point and found that $\dot{h}_+\approx sh_+$ with
$s$ being compatible with $p_{\rm M}/(\eta+1)$.
This suggests that the GW evolution is almost entirely dominated by the
rapid algebraic increase at each point in space.

It also turned out that, as expected from the work of
\cite{2017PhRvD..96h3511S}, the electric energy exceeds the
magnetic energy by a certain factor.
This factor depends on the value of $\beta$ and is about 8.6,
8.2, and 2.7 for $\beta=7.3$, 6.8, and 2.7, respectively.

\begin{figure*}\begin{center}
\includegraphics[width=\textwidth]{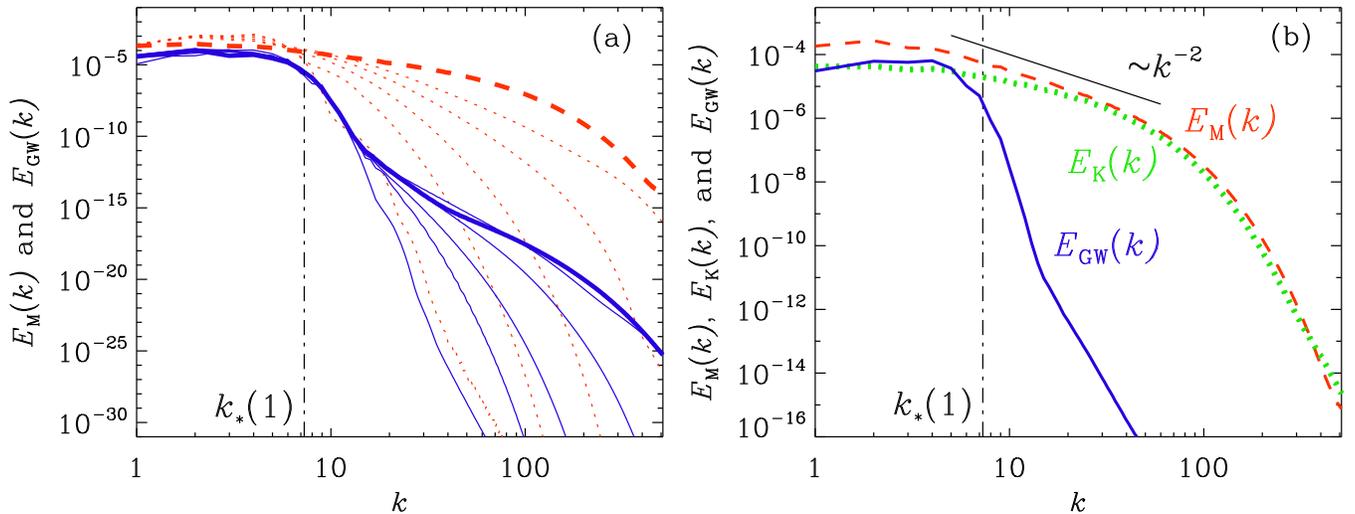}
\end{center}\caption{
(a) Early times in the beginning of the radiation-dominated phase:
$\eta=1$, $2$, $3$, $4$, $5$, and $20$ for Run~A1.
The last time at $\eta=20$ is shown as a thick red dashed line
for $\EM(k)$ and a thick blue solid line for $\EGW(k)$.
(b) Late times: $\eta=28$.
The vertical file goes through $k_*$ defined in \Eq{kpeak}.
}\label{rspec3_select_1024b68d_MHD}\end{figure*}

\begin{figure*}\begin{center}
\includegraphics[width=\textwidth]{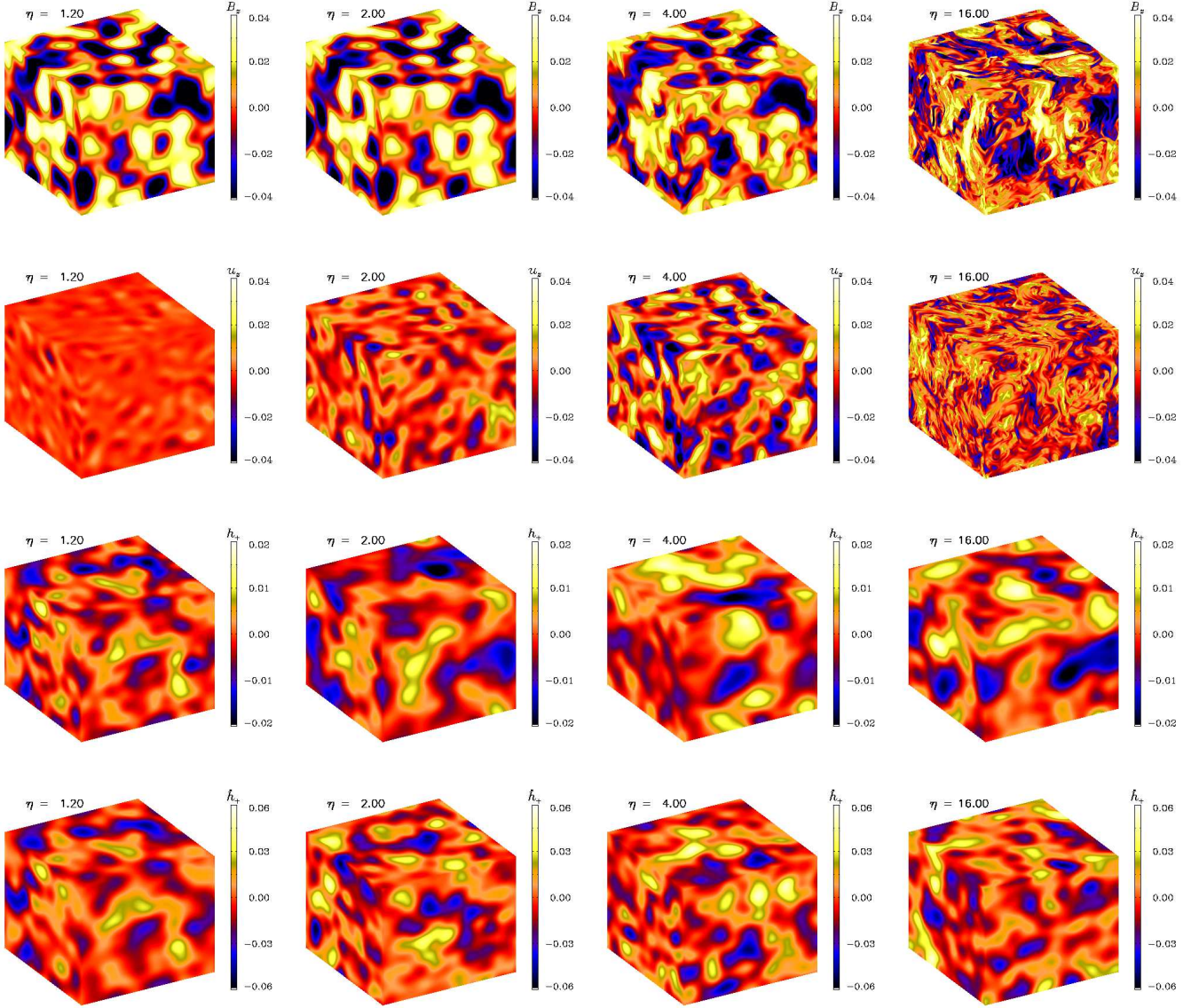}
\end{center}\caption{
Visualizations of $B_z$, $u_z$, $h_+$, and $\dot{h}_+$ (from top to bottom)
on the periphery of the computational domain for Run~A1 at four times
during step~II.
The color scale is different for each field, but the same at all times.
}\label{ABCD_1024b68d_MHD}\end{figure*}

\begin{figure*}\begin{center}
\includegraphics[width=.49\textwidth]{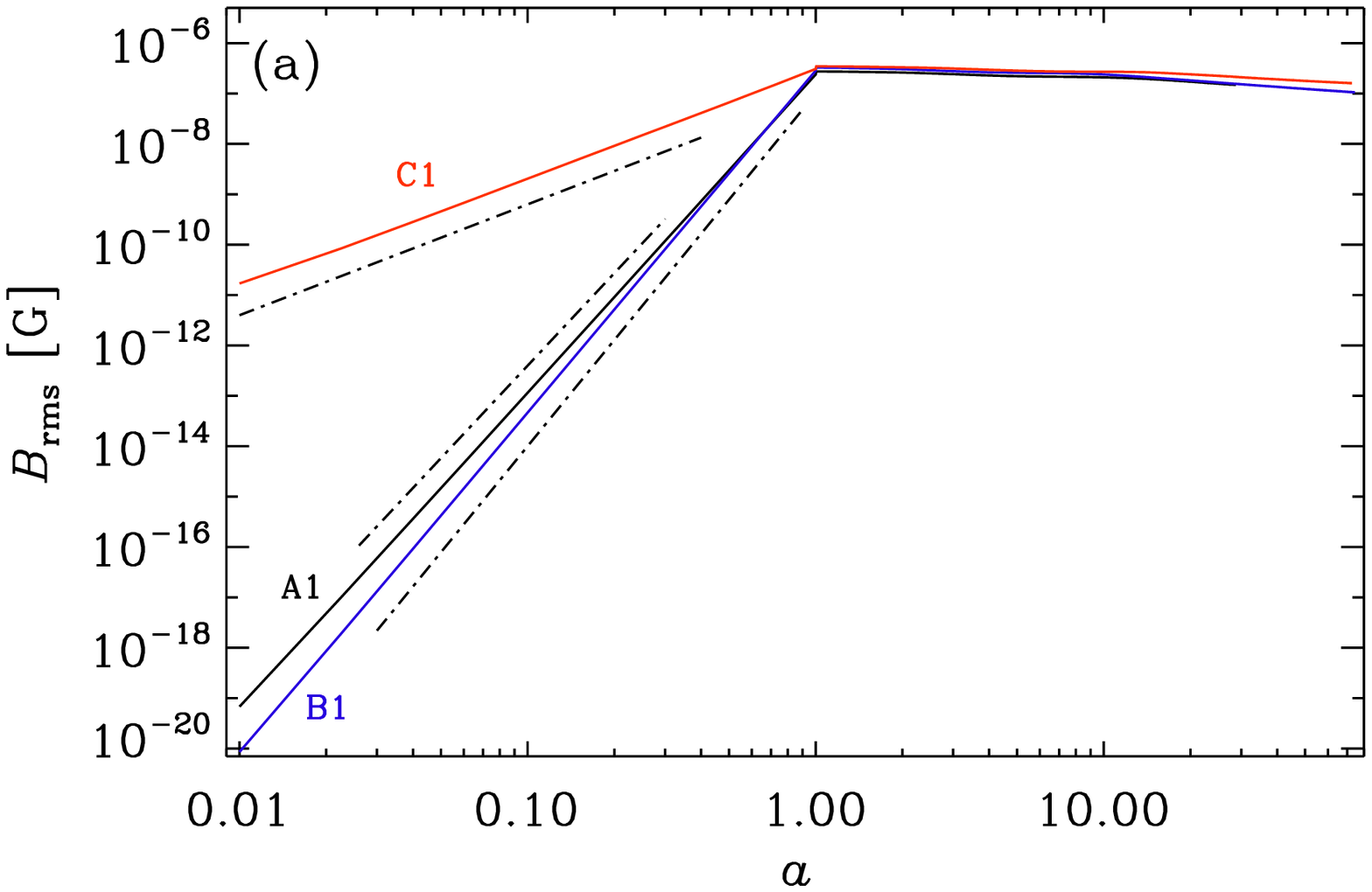}
\includegraphics[width=.49\textwidth]{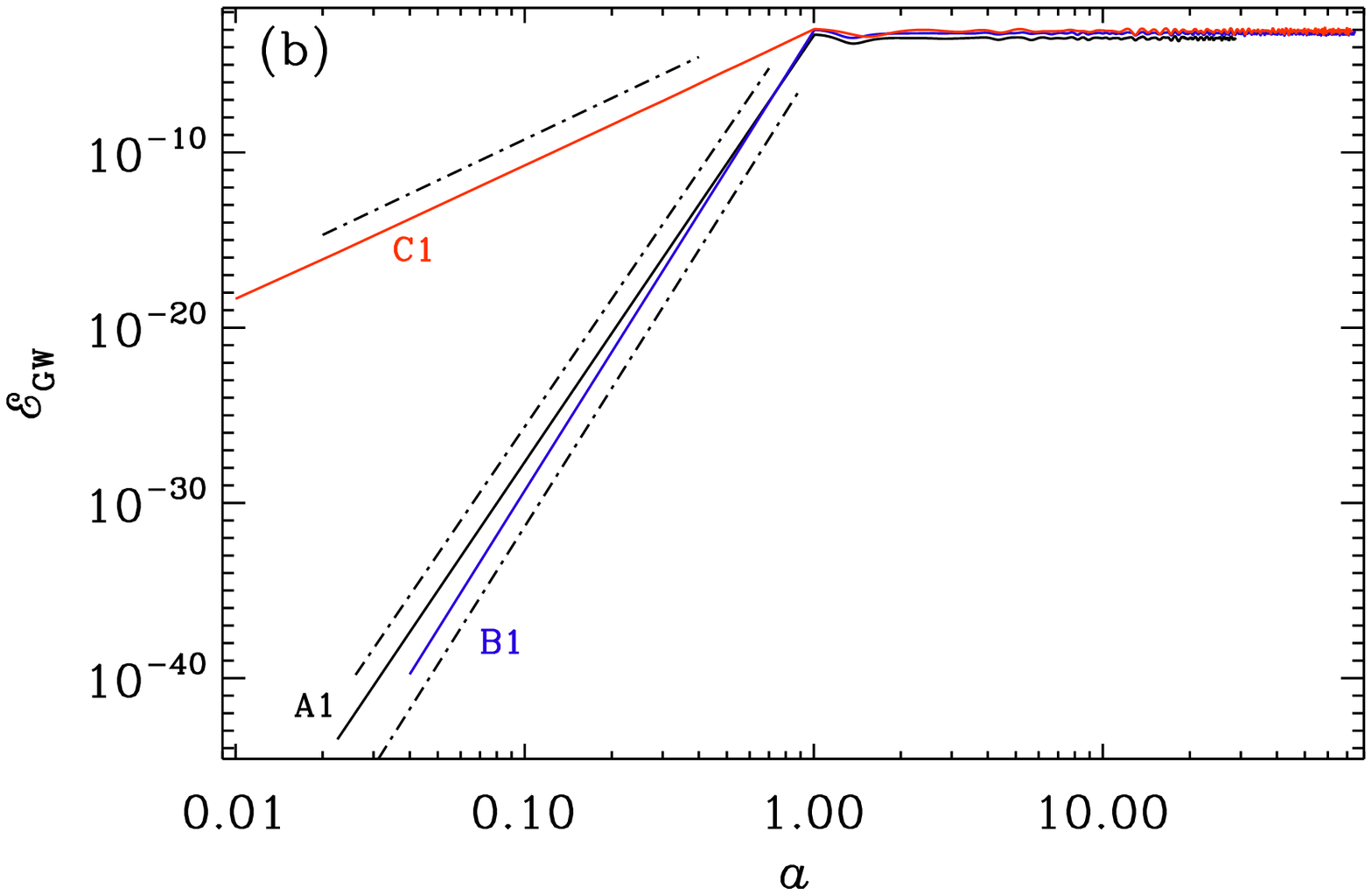}
\end{center}\caption{
Comparison of Runs~A1 (black line), B1 (blue line), and C1 (red line) showing
the evolution (expressed in terms of $a$) of 
(a) $\Brms$ (expressed in gauss), 
and (b) the $\EEGW$.
In (a), the dashed-dotted lines have slopes of 6.3, 6.8, and 2.2 for
Runs~A1, B1, and C1, respectively,
and in (b) the slopes are 24.2, 26.2, and 7.8.
}\label{pspect_comp_wav1}\end{figure*}

\begin{figure*}\begin{center}
\includegraphics[width=.49\textwidth]{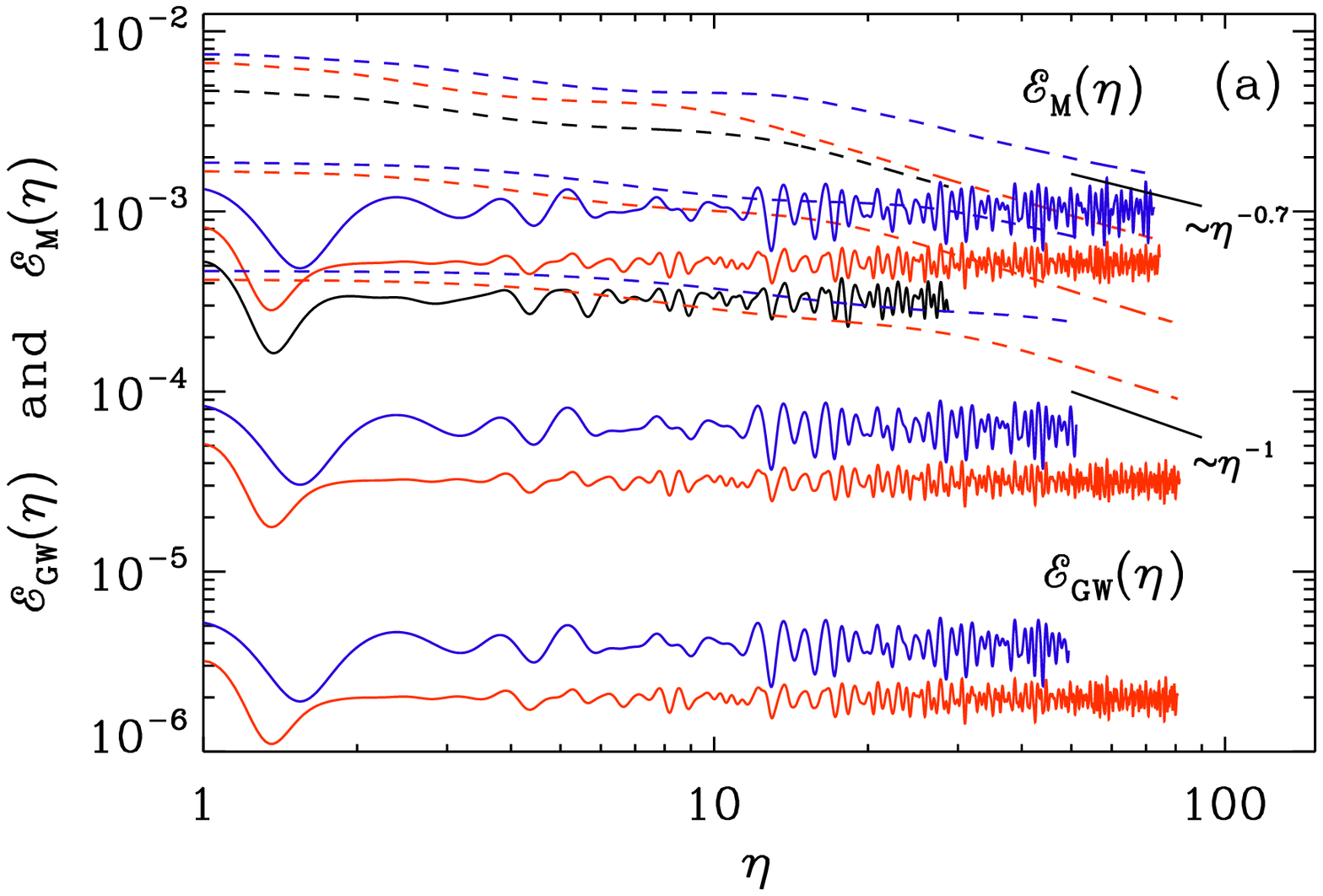}
\includegraphics[width=.49\textwidth]{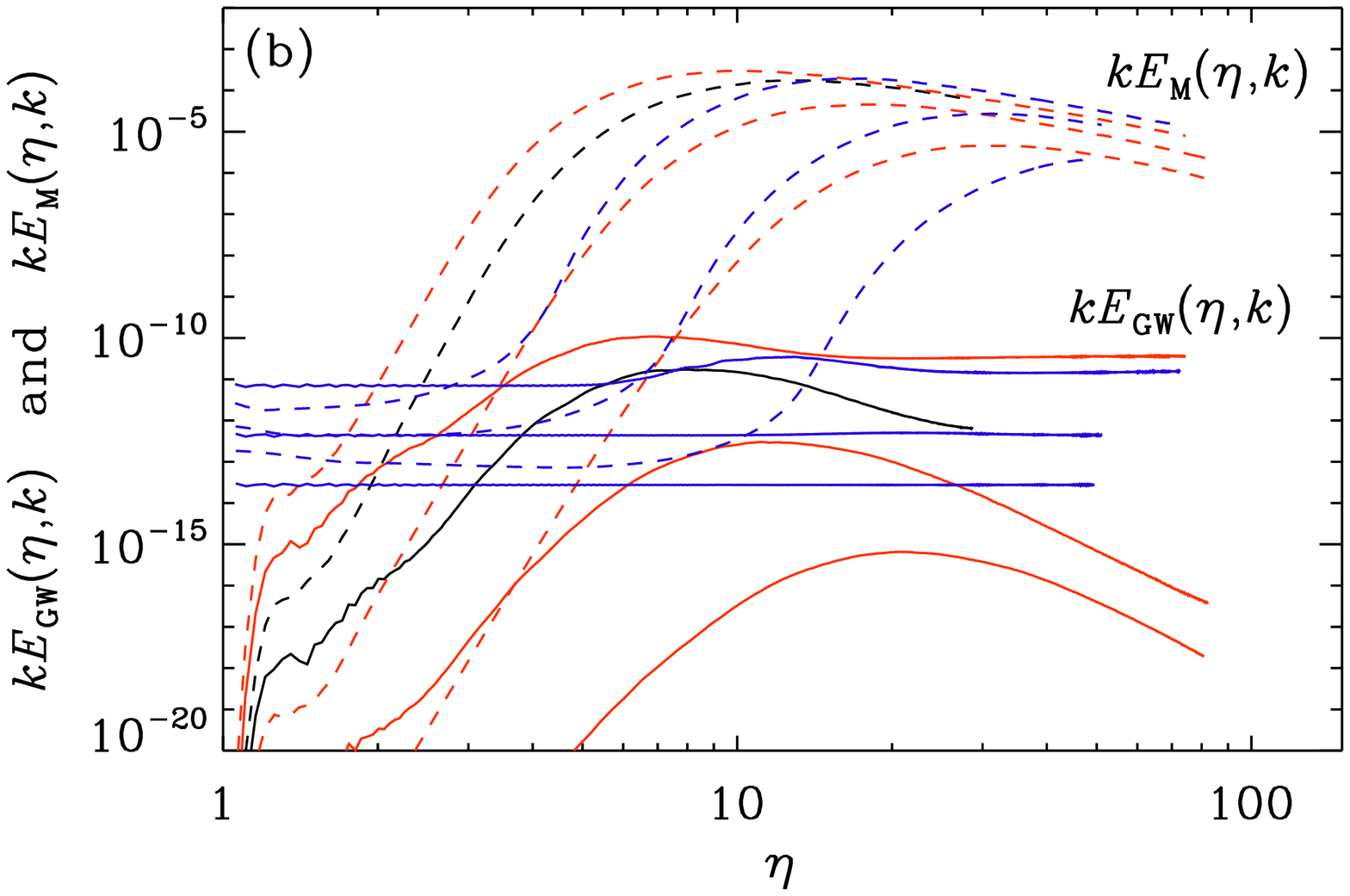}
\end{center}\caption{
Evolution of (a) $\EEGW$ and $\EEM$ as solid and dashed lines,
respectively, and (b) $k\EGW(\eta,k)$ and $k\EM(\eta,k)$, also as solid
and dashed lines, respectively, with $k=40$ for Runs~A1 (black line),
B1--B3 (blue lines), and C1--C3 (red lines).
In (a), the empirical decays $\propto\eta^{-0.7}$ and $\eta^{-1}$
are indicated.
}\label{pspect_comp2_wav1}\end{figure*}

Magnetic and electric fields still grow rapidly at the end of reheating,
but only at large length scales.
At $\eta=1$, we assume that the electric conductivity increases rapidly
to sufficiently high values, so there will be no electric fields
anymore, but there will be electric currents, $\JJ=\nab\times\BB$,
and they will exert a Lorentz force, $\JJ\times\BB$.
We then switch to MHD and solve for the resulting velocity field,
which facilitates a turbulent cascade toward smaller length scales.
In \App{ConductivityChanges}, we demonstrate quantitatively how
a faster increase of conductivity reduces the magnetic energy
loss during this transition into the high conductivity regime.

\subsection{Evolution during step~II}
\label{EvolStepII}

In all runs of step~II, we have initially $\uu=\ln\rho=0$.
We chose $\nu=10^{-4}$, which was the smallest possible value that
still allowed us to resolve the smallest length scales when $k^{-1}=1$
and $512^3$ mesh points were used.
For two pairs of runs (Runs~A1 and A2), we had to use $1024^3$ mesh points.
Yet smaller values of $\nu$ and the magnetic diffusivity $\sigma^{-1}=\nu$
would be physically more realistic, but would require an even larger
number of mesh points.
As is commonly known in turbulence theory, this would only extend the
turbulent cascade to smaller length scales, but would not strongly affect
the rest of the turbulent inertial range.

The magnetic, kinetic, and GW spectra are shown in
\Fig{rspec3_select_1024b68d_MHD}.
There is a gradual establishment of a turbulent cascade in magnetic
and kinetic energy spectra approximately proportional to $k^{-2}$.
However, during an intermediate stage of our investigations, we also have
experimented with even larger values of the exponent $\beta$ and found that
the turbulence in those cases is even more vigorous and can exhibit
a $k^{-5/3}$ spectrum, suggestive of Kolmogorov-like turbulence;
see \App{StrongerField} for an example.

The kinetic energy spectrum shows approximate equipartition at
small length scales, i.e.,
$\EK(k,\eta)\approx\EM(k,\eta)$ for $k>\kpeak(\eta)$.
The GW energy spectrum shows a characteristic drop at the smallest
unstable scale at the end of reheating, followed by
an approximate power-law spectrum at higher wavenumbers.
Such a drop was found particularly clearly in recent GW simulations
driven by an underlying magnetic field that was forced at very large
length scales \citep{2021arXiv210212428B}.
More generally, such a drop is seen to various extents in all GW
simulations sourced by monochromatically driven vortical turbulence;
see, for example, Figure~6 of \cite{2020GApFD.114..130R}.
By comparing with their Figure~4, one sees that this drop is not seen
when a turbulence spectrum is initialized through an
initial condition rather than through gradual driving.
This implies a sudden jump in time, even at small length scales,
which is unrealistic.
Our simulations predict for the first time a natural time scale of
the temporal increase of the stress, especially at high wavenumbers.
The relatively low amplitude of GWs at these high wavenumbers suggests
that GW generation is predominantly a large-scale phenomenon and therefore
also not strongly dependent on the exact details at small length scales.
Another reason for this sharp drop at higher $k$ is that the turbulent
stress develops only later, when the $1/a$ factor on the right-hand side
of \Eq{d2hdt2} has diminished its effect.

We find the ratio between magnetic and kinetic energies to be only about
1.3 and not as large as in some earlier simulations of turbulence driven
by an initial magnetic field with a spectrum peaked at intermediate
wavenumbers, where $\EM\propto k^{-2}$ was found.
In the present case, the spectrum is peaked at large length scales,
so there is no possibility for the magnetic field to display marked
inverse transfer to larger length scales, as was the case in simulations
of \cite{2015PhRvL.114g5001B}, who found inverse transfer even without
magnetic helicity; see also \cite{2014ApJ...794L..26Z} for relativistic
turbulence simulations.

Earlier work on inflationary magnetogenesis presumed the appearance
of a scale-invariant spectrum proportional to $k^{-1}$; see, e.g.,
\cite{2012PhRvD..86j3005K,2017JCAP...12..002K}.
In the present case, the magnetic field along with velocity fluctuations
are being driven by a turbulent cascade and fed by the large-scale
magnetic field.
As already noted by \cite{2017PhRvD..96h3511S,2018PhRvD..97h3503S},
the magnetic field has a blue $k^3$ spectrum at small wavenumbers, $k<k_*(1)$.
During the early part of the radiation-dominated phase, we see that the peak
gradually shifts to smaller wavenumbers, so the initial $k^3$ spectrum can
hardly be recognized within the limited wavenumber range accessible
to our simulations; see \Figp{rspec3_select_1024b68d_MHD}{b}.

\begin{figure*}\begin{center}
\includegraphics[width=\textwidth]{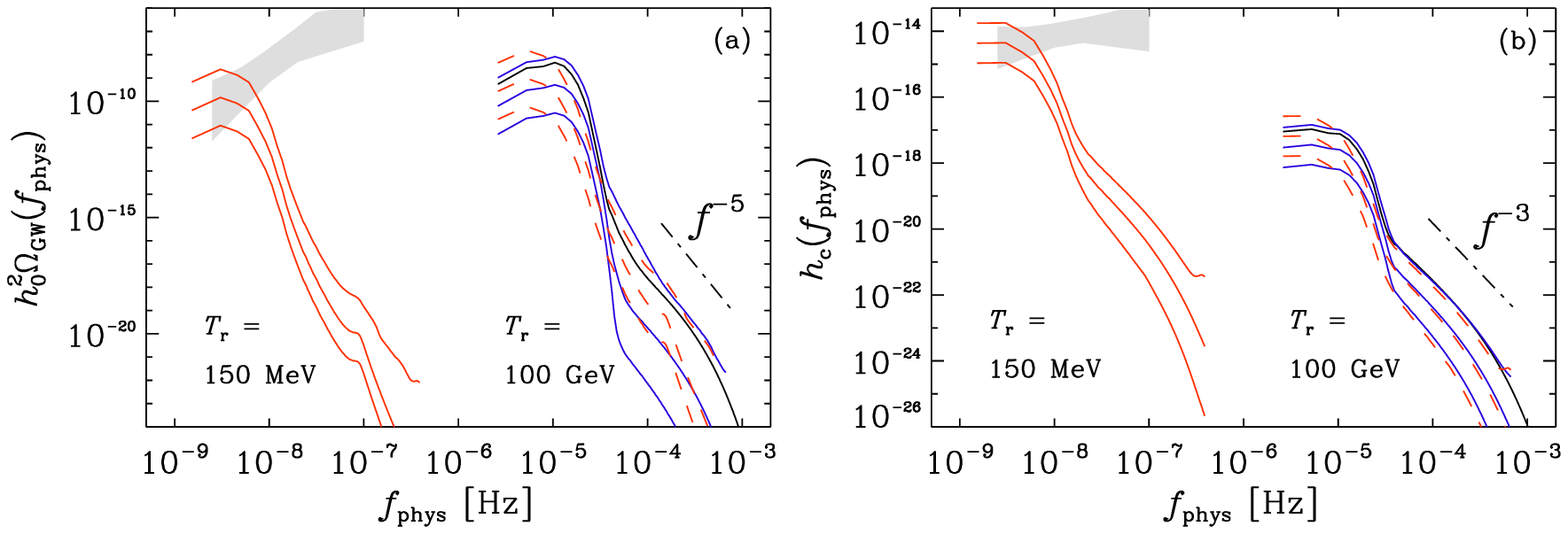}
\end{center}\caption{
(a) $h_0^2\Omega_{\rm GW}(f_{\rm phys})$ and (b) $h_c(f_{\rm phys})$ for Runs~A1 (black lines)
and B1--B3 (blue lines) for $T_{\rm r}=100\GeV$ and Runs~C1--C3 (red lines)
for $T_{\rm r}=150\MeV$.
To compare the two energy scales, we show Runs~C1--C3 also for
$T_{\rm r}=100\GeV$ (red dashed lines).
The gray region marks the $2\sigma$ confidence contour for the
30 frequency power law of the NANOGrav 12.5-year data set.
}\label{pspecm}\end{figure*}

In \Fig{ABCD_1024b68d_MHD}, we show visualizations of $B_z$, $u_z$,
$h_+$, and $\dot{h}_+$ on the periphery of the computational domain
for Run~A1 during step~II at $\eta=1.2$, 2, 4, and 16.
We see that for $\eta\leq2$, the magnetic field has almost not
changed at all.
The velocity is still small, but begins to become important for $\eta>2$.
Fully developed turbulence is seen at $\eta=16$.
However, the strain field and its time derivative are not visibly
affected by the fully developed turbulence.

\subsection{Time series for different values of $\beta$}
\label{DifferentBeta}

In \Fig{pspect_comp_wav1}, we show the evolution of $\Brms$ and $\EEGW$
both for steps~I and II as a double-logarithmic plot.
Since $\eta$ can be negative, we express time in terms of $a=(\eta+1)^2/4$
for $\eta<1$ (and $a=\eta$ otherwise).
Owing to the quadratic scaling in time and the additional quadratic
scaling of magnetic energy with $\Brms$ in step~I, the slopes of $6.8$,
$6.3$, and $2.2$ for Runs~A1--C1 correspond to the exponents of
$p_{\rm M}=27.2$, $25.2$, and $8.8$, respectively.
In step~II, we see that $\Brms$ displays a comparatively slow decay
relative to the rapid increase for $\eta<1$.
The decay for $\eta\gg1$ follows a power law
$\propto\eta^{-1}$ for $\beta=7.3$, and $\propto\eta^{-0.7}$
for $\beta=2.3$; see \Figp{pspect_comp2_wav1}{a}.
This panel also shows that the GW energy fluctuates in time, but is
otherwise statistically stationary.
There is, however, a systematic wiggle in all curves of $\EEGW$ at
around $\eta=1.05$.
This is caused by the discontinuity in $a''/a$ and $f''/f$ at $a=1$.
In \App{Avoiding}, we examine the effects of removing the
discontinuity on the occurrence of oscillations and we
also study the effects on the GW energy spectrum between
the end of step~I and the beginning of step~II.

A decay of $\EEM$ proportional to $\eta$ is also what has been obtained in
earlier simulations of magnetically dominated decaying turbulence
\citep{2017PhRvL.118e5102B}, but the slower decay proportional to
$\eta^{-0.7}$ has only been seen in the presence of magnetic helicity.
Here, however, the magnetic helicity is zero.
The reason for this slower decay is probably connected with the absence
of an extended subinertial range in our simulations, where $\kpeak(1)$
is too close to the minimal wavenumber $k_1$.
If we allowed for more mesh points and larger domains, the expected
$\eta^{-1}$ decay should be recovered.

Our simulations yield a temporal increase of the GW energy at
length scales smaller than $k_*^{-1}(\eta)$ for $\eta>1$.
In the absence of turbulence, GWs would only have existed on large
length scales.
To see the development at intermediate length scales more clearly, we
compare in \Fig{pspect_comp2_wav1} the temporal evolution of
the GW and magnetic energy densities in panel (a), and in
panel (b) the GW and magnetic energies at the wavenumber $k=40$.
We see that magnetic and GW energies increase with time.
The larger the magnetic energy at $\eta=1$, the more rapidly $\EEGW(\eta)$
increases and the larger is their final GW energy.
There is considerable spread in the final values of the scale-dependent GW
energy densities, while the spread in the scale-dependent {\em magnetic}
energy densities is much less.
Unlike the total, wavenumber-integrated GW energy, which is nearly
perfectly statistically stationary already after a short time, the
scale-dependent values are in some cases not yet steady and are still
decreasing after having reached a certain maximum value.

\begin{table}[b]\caption{
$C_1$, $C_2$, and $C_*$ for two values of $T_{\rm r}$.
}\begin{center}
\begin{tabular}{cccc}
$T_{\rm r}$ & $H_*/H_0=C_1$ & $a_*/a_0=C_2$ & $C_*=C_1^2C_2^4$ \\
\hline
$100\GeV$ & $6.4\times10^{27}$ & $8.0\times10^{-16}$ & $1.6\times10^{-5}$ \\
$150\MeV$ & $5.6\times10^{21}$ & $1.0\times10^{-12}$ & $3.1\times10^{-5}$ \\
\label{Tconversion}\end{tabular}
\end{center}
\end{table}

\begin{table*}\caption{
Summary of simulation parameters and properties.
}\begin{center}
\begin{tabular}{ccc|cc|cc|c|c|cc}
\multicolumn{5}{c}{} &
\multicolumn{2}{c|}{radiation dominated} & $\Brms$ &
\multicolumn{1}{c}{$T_{\rm r}$} &
\multicolumn{2}{c}{scaled to the present time} \\
Run & $B_0$ & $\beta$ & $\EEEM$ & $\EEEl/\EEM$ & $\EEGW$ & $\hrms$ & [$\mu$G] & [GeV] & $\OmGW$ & $h_{\rm c}$ \\
\hline
A1&$1\times10^{-17}$&$6.8$&$0.12$&$32.7$&$3.3\times10^{-4}$&$1.7\times10^{-2}$&$0.24$&$100$&$5.4\times10^{-9}$&$1.4\times10^{-17}$\\
A2&$1\times10^{-17}$&$6.8$&$0.12$&$32.7$&$3.2\times10^{-4}$&$1.7\times10^{-2}$&$0.24$&$100$&$5.3\times10^{-9}$&$1.4\times10^{-17}$\\
A3&$1\times10^{-17}$&$6.8$&$0.12$&$32.7$&$5.9\times10^{-5}$&$8.5\times10^{-3}$&$0.24$&$100$&$9.7\times10^{-10}$&$6.8\times10^{-18}$\\
B1&$1\times10^{-18}$&$7.3$&$0.18$&$34.4$&$6.3\times10^{-4}$&$2.3\times10^{-2}$&$0.28$&$100$&$1.0\times10^{-8}$&$1.9\times10^{-17}$\\
B2&$5\times10^{-19}$&$7.3$&$0.05$&$34.4$&$3.9\times10^{-5}$&$5.8\times10^{-3}$&$0.14$&$100$&$6.3\times10^{-10}$&$4.6\times10^{-18}$\\
B3&$2\times10^{-19}$&$7.3$&$0.01$&$34.4$&$2.4\times10^{-6}$&$1.4\times10^{-3}$&$0.07$&$100$&$3.9\times10^{-11}$&$1.1\times10^{-18}$\\
C1&$5\times10^{-7}$&$2.7$&$0.07$&$10.8$&$8.6\times10^{-4}$&$3.7\times10^{-2}$&$0.31$&$0.15$&$2.7\times10^{-8}$&$3.7\times10^{-14}$\\
C2&$2\times10^{-7}$&$2.7$&$0.017$&$10.8$&$5.3\times10^{-5}$&$9.1\times10^{-3}$&$0.15$&$0.15$&$1.6\times10^{-9}$&$9.1\times10^{-15}$\\
C3&$1\times10^{-7}$&$2.7$&$0.004$&$10.8$&$3.3\times10^{-6}$&$2.3\times10^{-3}$&$0.08$&$0.15$&$1.0\times10^{-10}$&$2.3\times10^{-15}$\\
\label{Tsummary}\end{tabular}
\end{center}
\end{table*}

\subsection{Present-day frequency spectra}

To compute the strain at the present time, we have to multiply
our values of $\hrms$ by the ratio $C_1=a_*/a_0$, where $a_*$
and $a_0$ are the scale factors at reheating and the present time, respectively.
To obtain the GW energy at the present time, we also have to take the
Hubble factor $C_2=H_*/H_0$ into account, where $H_*$ and $H_0$
are the Hubble parameters at reheating and the present time, respectively.
Thus, we have to multiply our value of $\EEGW$ by the dilution
factor, $C_*\equiv(a_*/a_0)^4(H_*/H_0)^2\equiv C_1^4 C_2^2$; see
\cite{2020PhRvD.102h3512R}.
The values of $C_1$, $C_2$, and $C_*$ are given in
\Tab{Tconversion}, and are also consistent with those used
earlier \citep{2021arXiv210212428B,2021arXiv210403192H}.

The resulting GW energy and strain spectra, scaled to the present
time, are shown in \Fig{pspecm}, where we show the frequency spectra
\EQ
h_0^2\Omega_{\rm GW}(f_{\rm phys})=C_*k\EGW(k); \; h_c(f_{\rm phys})=C_2\hrms(k)
\EN
with $f_{\rm phys}(k)=H_*k/2\pi a_0$ being the physical frequency
for Runs~A1--C3; see also \Tab{Tsummary} for the parameters.
In Run~A3, the GW field has been reset to zero at $\eta=1$ (see
\Sec{SignificanceStepI}), but in all other cases, $\tildeh_{+/\times}$
and $\tildeh'_{+/\times}$ have been inherited from step~I.
In \Fig{pspecm}, we have also indicated the $2\sigma$ confidence
contour for the 30 frequency power law of the NANOGrav 12.5-year data
set \citep{NANOGrav2020}; see \cite{2021arXiv210212428B} for the details
of the determination of the present contours.

Runs~B1--B3 have the same value of $\beta$, but different
initial magnetic field strengths, resulting also
in different values of the maxima of $\EEEl$ and $\EEM$.
We recall that these values are automatically the fractional magnetic
energy density relative to the radiative energy density, which is also
equal to the critical energy density for a spatially flat Universe.
Our choice of $\beta=7.3$ applies specifically to the case with
$\EEEM\equiv\EEEl+\EEM=0.01$, which corresponds to the case of Run~B3.
For Run~B1, on the other hand, where $\EEEM$ is larger, the value
$\beta=6.8$ should have been more appropriate.
Such a case is presented in \Tab{Tsummary} as Run~A1.
However, we see that the differences between Runs~A1 and B1 are minor.
We can therefore conclude that the precise choice of $\beta$ is
not critical for the final outcome of our models.

\Fig{pspecm} shows that, even well beyond the drop in
$h_0^2\Omega_{\rm GW}(f_{\rm phys})$, there continues to be a fairly
steep fall-off proportional to $f^{-5}$.
Such a strong decline was also seen in the earlier work of
\cite{2020PhRvD.102h3512R}; see their Figure~6.
This is much steeper than the $f^{-8/3}$ law obtained for the
GW spectrum when a turbulent source with a Kolmogorov-type spectrum
is switched on as the initial condition.
In \App{StrongerField}, we demonstrate that such a shallow fall-off can be
reproduced by assuming an artificially strong source where the magnetic
energy exceeds 10\% of the radiation energy density and therefore the
electric field energy density just prior to the commencement of MHD
would have been as large as the radiation energy density of the universe.

\begin{figure*}\begin{center}
\includegraphics[width=\textwidth]{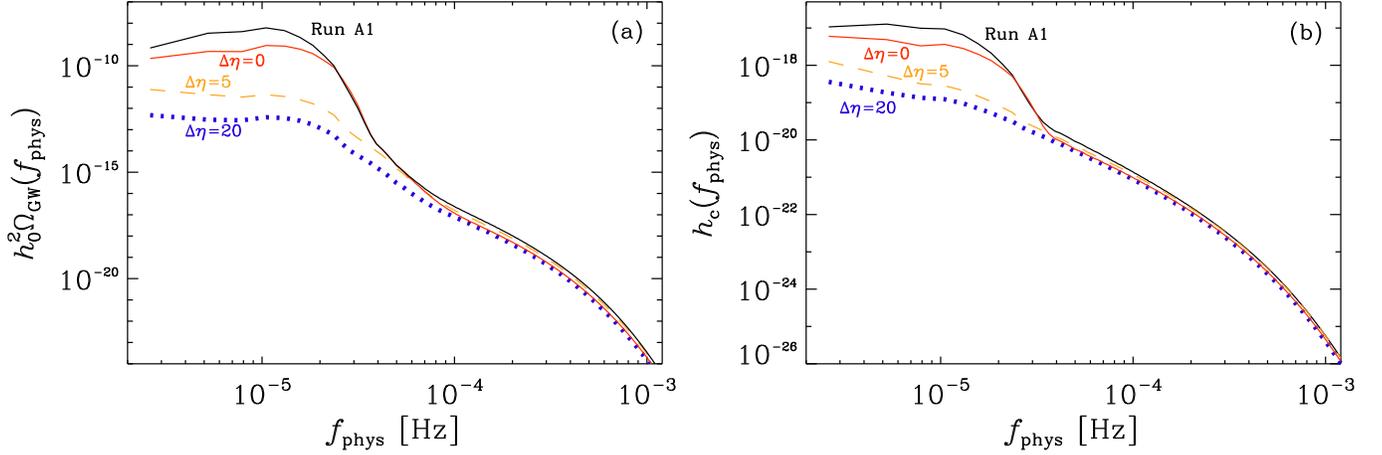}
\end{center}\caption{
(a) $h_0^2\Omega_{\rm GW}(f_{\rm phys})$ and (b) $h_c(f_{\rm phys})$
for Run~A1 compared with runs where the GW field from step~I has been
ignored, so $\tildeh_{+/\times}=\tildeh'_{+/\times}=0$ has been set
and the hydromagnetic stress was applied instantaneously (Run~A3,
$\Delta\eta=0$, red solid line), or gradually over a time span
$\Delta\eta=5$ (orange dashed line) or $\Delta\eta=20$ (blue dotted line).
}\label{pspecm_ramp}\end{figure*}

Earlier work by \cite{2020PhRvD.102h3512R} did already demonstrate
that the Reynolds stress in $\TTT_{ij}$ contributes only about 10\%
to the resulting GW energy.
In \Tab{Tsummary}, we have listed Run~A2, where the Reynolds
stress is omitted in $\TTT_{ij}$.
It turns out that the GW energy is indeed reduced, but only by a
few percent, so the Reynolds stress is here completely unimportant.
This is partially explained by the fact that the resulting turbulence
develops almost entirely on small scales (see \Fig{ABCD_1024b68d_MHD}),
where the effect on GWs is small.

We see that most of the power occurs at frequencies of about $20\uHz$
($10\nHz$) for $T_{\rm r}=100\GeV$ ($150\MeV$), followed by a drop of
GW energy and strain by four and two orders of magnitude, respectively.
This may suggest that the turbulent production of GWs is only moderately
effective in converting the turbulent energy from the forward cascade
to GW energy.
In the following, we shall look at this more quantitatively.

\begin{figure*}\begin{center}
\includegraphics[width=\textwidth]{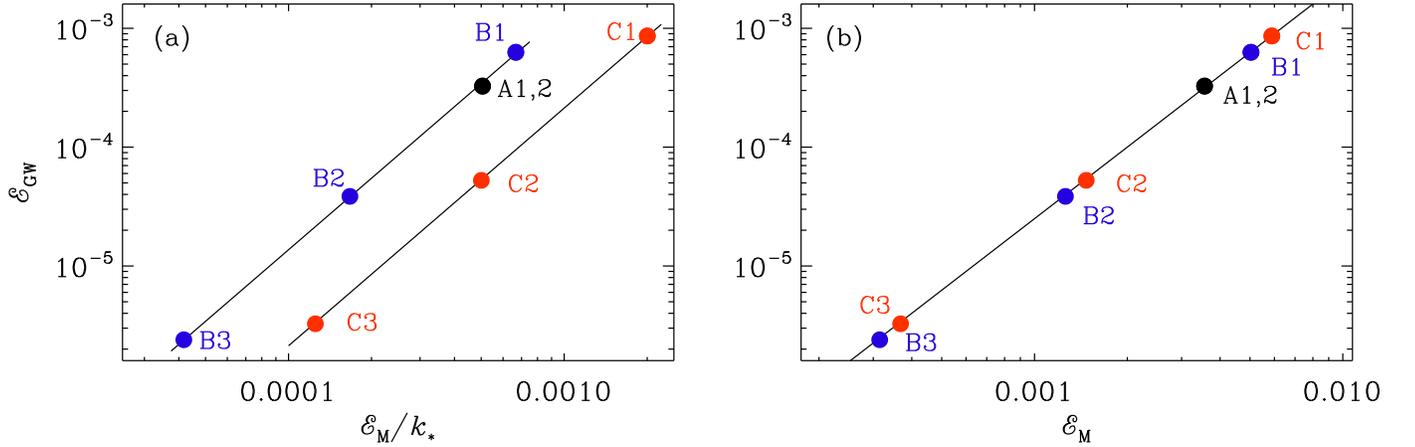}
\end{center}\caption{
(a) Dependence of $\EEGW$ on $\EEM/k_*$ for Runs~A1 and A2 (black),
B1--B3 (blue), and C1--C3 (red).
The solid line is a quadratic fit through the black and blue symbols,
$\EEGW=(q\EEM/k_*)^2$, with $q=37$
and through the red symbols with $q=14.6$.
(b) Similar to panel (a), but plotted versus $\EEM$.
Here, the solid line corresponds to the fit given by \Eq{fit}.
}\label{ptab2}\end{figure*}

\subsection{Significance of using the GW field from step~I}
\label{SignificanceStepI}

Except for the recent simulations of \cite{2021ApJ...911..110B}, previous
work on numerical investigations of GW generation from hydrodynamic or MHD
turbulence assumed that turbulence was either switched on instantaneously
or it was gradually being produced; see \cite{2020PhRvD.102h3512R}
for comparisons of such models.
In either case, the GW field was always initially zero, i.e.,
$\tildeh_{+/\times}=\tildeh'_{+/\times}=0$.
Here, by contrast, both $\tildeh_{+/\times}$ and
$\tildeh'_{+/\times}$ are finite at $\eta=1$.
As suggested in the introduction, this can make a difference and
could underestimate the resulting GW energy.
To study this in more detail, we now perform an additional simulation
(Run~A3 in \Tab{Tsummary}) with $\tildeh_{+/\times}=\tildeh'_{+/\times}=0$
at $\eta=1$, using just the magnetic field from step~I.
We see that the GW energy is now about 10 times weaker than otherwise
(Run~A1).

The resulting spectra are shown in \Fig{pspecm_ramp}, where we plot
the GW energy and strain for a series of models where we turned on
the stress either instantaneously (as in Run~A3) or gradually using a
linearly varying profile function that multiplies the stress by a factor
that linearly grows to unity within a time to span $\Delta\eta$.
Thus, we replace
\EQ
\tildeT_{+/\times}\to\frac{\eta-1}{\Delta\eta} \tildeT_{+/\times}
\quad\mbox{for $1<\eta<1+\Delta\eta$}.
\EN
The case of instantaneously switching on the stress corresponds
then to $\Delta\eta\to0$.
The original Run~A1 is shown for comparison.

We see that an instantaneously switched on stress produces a GW spectrum
that agrees with the original one at high wavenumbers, and only at low
wavenumbers is there a small deficiency in GW energy and strain.
This shows that the inheritance of the GW field from step~I is of
relatively minor importance.
The agreement at high wavenumbers is no surprise because at those
high $k$ values, the GW field was absent at $\eta=1$.
The speed of regeneration of the GW field at low $k$ is more surprising,
but the qualitative agreement with Run~A1 is probably related to the fact
that the generation in step~I is so rapid that only the last moment has
a decisive effect on the GW field.
This then also demonstrates that the reason for the rapid drop in spectral
GW energy and strain past the peak wavenumber is, at least in part, a
consequence of the existence of magnetic fields only for $k<\kpeak(1)$
at $\eta=1$.

To examine this further, we now describe two models where $\Delta\eta=5$
and $20$.
We see that the dominance of the GW energy and strain at small $k$
diminishes and that also the sharp drop in spectral GW energy and strain
becomes smaller.
Whether the remaining lack of continuity of the GW spectrum at
$k\approx\kpeak(1)$ is caused by the limited vigor of turbulence
is unclear.
Nevertheless, the existence of the sharp drop in spectral GW energy and
strain seems to be a physical effect that was not previously
anticipated in relic GW modeling.

\begin{figure*}\begin{center}
\includegraphics[width=\textwidth]{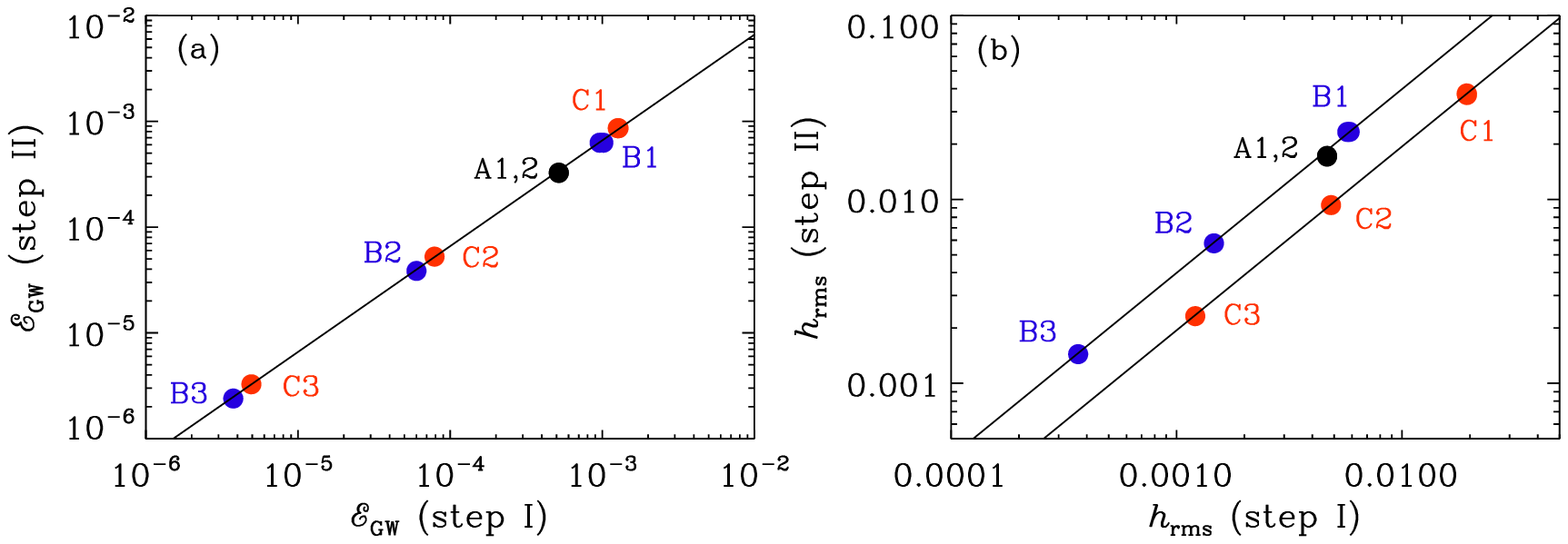}
\end{center}\caption{
Dependence of the final values in step~II of (a) $\EEGW$ and 
(b) $\hrms$ on those at the end of step~I.
In (a) the solid line denotes $\EEGW(\mbox{step~II})=0.66\,\EEGW(\mbox{step~I})$, while
in (b) the upper solid line denotes $\hrms(\mbox{step II})=4.0\,\hrms(\mbox{step I})$
for $\beta=7.3$ and the lower solid line denotes
$\hrms(\mbox{step~II})=1.9\,\hrms(\mbox{step~I})$ for $\beta=2.7$.
}\label{pEEGW_step12}\end{figure*}

\subsection{GW efficiency and scaling with $\EEM$}

The GW energy of our runs scales approximately quadratically with magnetic energy.
Following earlier work \citep{2020PhRvD.102h3512R,2021arXiv210301140B},
we confirm a relation of the form $\EEGW=(q\EEM/k_{\rm c})^2$, where
$q$ is the efficiency and $k_{\rm c}$ is the characteristic wavenumber,
for which the value $k_{\rm c}=\kpeak(1)$ has been used.
The values of $\EEM$ range between 0.03\% and 0.5\% of the radiation
energy density.

With that, we find for Runs~B1--B3 an efficiency parameter of $q=37$,
and smaller values of $14.6$ for Runs~C1--C3, where $\beta=2.7$; see \Figp{ptab2}{a}.
The obtained efficiency $q$ is smaller for smaller values of $\beta$,
suggesting a dependence between $q$ and $\beta$.
In particular, using $q=5\beta$ appears to be a good empirical
description of our data.
However, since $\kpeak$ also depends on $\beta$, and the ratio
$\beta/\kpeak(1)$ is about unity, a good fit to the data is then given by
\EQ
\EEGW\approx\,(5\,\EEM)^2.
\label{fit}
\EN
We thus see that the previously obtained $k_{\rm c}$ dependence of $\EEGW$
\citep{2020PhRvD.102h3512R, 2021arXiv210212428B, 2021arXiv210301140B}
can just be subsumed into the dependence on $\EEM$, at least in the
present case.

\subsection{Change of GW amplitude between steps~I and II}

It turns out that the final values of $\EEGW$ and $\hrms$ are not the
same as those at the end of step~I.
However, they are proportional to each other in such a way that the
final $\EEGW$ in step~II is about 0.66 times the value at the end
of step~I, and the final $\hrms$ in step~II is about four times the
value at the end of step~I for $\beta=7.3$ and about twice the
value at the end of step~I for $\beta=2.7$; see \Fig{pEEGW_step12}.

\section{Conclusions}

We have presented three-dimensional direct numerical
simulations of inflationary magnetogenesis and relic GW production during
the end of a matter-dominated reheating era, and the subsequent evolution
in the beginning of the radiation-dominated era.
As expected based on earlier analytic work, electromagnetic fields grow in
power-law fashion, with the electric field exceeding the magnetic one.
GWs are driven by the electromagnetic stress and also grow in power-law fashion.
The growth terminates with the beginning of the radiation-dominated era,
when high electric conductivity leads to a turbulent MHD cascade.
Vigorous motions produce small-scale hydromagnetic stresses, but they
are too weak to have a significant effect on the GW spectrum, which
therefore remains being dominated by large-scale features.
This is seen as a marked drop in the GW energy at present-day frequencies
of around $20\uHz$ ($3\nHz$) for a reheating temperature of $100\GeV$
($150\MeV$).

In comparison with earlier analytical work estimating the efficiency of GW
production from inflationary magnetogenesis by \cite{2020PhRvD.101j3526S},
our present work has highlighted some important aspects and discrepancies
with numerical modeling.
There is, most importantly, the spectral drop in the GW spectra above
the peak wavenumber $\kpeak(1)$, which remained almost unchanged after
$\eta=1$.
The drop is clearly seen both in wavenumber spectra
(\Fig{rspec3_select_1024b68d_MHD}) and in the diagnostic
frequency spectra (\Fig{pspecm}).
Here, the frequency spectra have been obtained from the wavenumber
spectra, but let us emphasize in this connection that the numerical
equivalence between the two was recently confirmed and that even a
spectral drop similar to that seen here is faithful reproduced from the
temporal Fourier transform of the time series \citep{2021arXiv210403192H}.
Discrepancies between temporal and spatial spectra can occur, however,
when the dispersion relation of GWs is no longer linear (for example
for finite graviton mass), or when there is a long-term effect from the
stress associated with the slowly decaying turbulence
\citep{2021arXiv210403192H}.
This effect only plays a very small role because most of the GW energies
are dominated by contributions from small wavenumbers.
Nevertheless, it could explain small discrepancies between strain spectra
and the anticipated GW energy spectra at frequencies above the spectral
drop of the GW energy.
This drop can then reveal subtle features associated with the turbulent
inertial range, but the aforementioned differences should disappear at
very late times, well beyond what has been simulated so far, and would
not be observationally significant.

The spectral drop in GW energy is not a particular feature associated with
inflationary magnetogenesis, but it appears to be a feature associated with
turbulence driven at scales comparable to the horizon scale with $k\approx1$.
It should be emphasized that, at the level of the present model, there
is no immediate association between the value of the reheating temperature
and the physics of phase transitions.
In particular, there is no obvious feature in the resulting GW spectra
from magnetogenesis during reheating and any other hypothetical source
of turbulence.
This can be seen by comparing the present GW spectra with those of
\cite{2021arXiv210212428B}, where turbulence was driven by an assumed
low wavenumber forcing function.
A possible difference may lie in the dependence of $\EEGW$ on $\EEM$,
which does not involve an independent scaling with the inverse of the
characteristic wavenumber $k_{\rm c}$.
This finding came as somewhat of a surprise, but it may just mean that,
in the present model, the efficiency parameter does indeed scale with
$\beta$.
Preliminary experiments with helical magnetogenesis \citep{Bran+He+Shar21}
suggest that the $1/k_{\rm c}$ scaling remains in general justified.

A particular problem in previous numerical modeling of GW production by
magnetic stresses laid in the fact that most of the GW energy resides at
small wavenumbers or large length scales.
This was particularly evident in some of the simulations describing
the chiral magnetic effect; see Run~B1 of \cite{2021ApJ...911..110B}
for such an example.
This meant that such results remained sensitive to the value of $k_1$
and thus the choice of the size of the computational domain.
In the present work, we have alleviated this problem by extending our
domain to larger length scales, so that the spectral peak was still well
within the domain of the model.
The strength of the underlying magnetic field was then only limited
by the condition that the electromagnetic energy should not exceed
about 10\% of the radiation energy density at the end of reheating.
Owing to the subsequent emergence of a turbulent cascade, the magnetic
energy is then being fed into smaller length scales.
This leads to a temporal growth of both magnetic and GW energies at
subhorizon length scales.
However, the strength of GWs at small length scales remained weak
compared with that at larger length scales.
We have not yet seen a strong dependence on the numerical resolution.
In fact, while both velocity and magnetic fields showed a well-resolved
fine structure, the strain field and also its time derivative remained
dominated by large-scale features.
It is conceivable that even higher numerical resolution would be needed to
produce sufficiently rapid variations at small length scales to enhance
it GW production at those small scales, but at the moment there is no
evidence for this.

An important extension of the present work is to consider helical
magnetogenesis \citep{1988PhRvD..37.2743T, 1992PhRvD..46.5346G,
2000PhRvD..62j3008F, 2006JCAP...10..018A, campanelli2008,
2014JCAP...10..056C,adshead2016,sobol2019,adshead2020a,adshead2020b}.
These authors considered an additional chiral symmetry breaking term
involving the dual Faraday tensor in the Lagrangian; see also the papers
by \cite{2018PhRvD..97h3503S} and \cite{2021JCAP...03..026O}.
Helical magnetic fields decay more slowly than nonhelical ones and are
therefore more likely to survive until the present time; see Figure~11
of \cite{2017PhRvD..96l3528B}.
It would then also be interesting to see whether there are any other
similarities with magnetogenesis from the chiral magnetic effect
\citep{2021ApJ...911..110B}, in addition to the relation between the GW
efficiency and the exponent $\beta$ identified in the present work.

\begin{acknowledgements}
We thank Tina Kahniashvili and Kandaswamy Subramanian for inspiring
discussions.
We also thank the anonymous referee for making useful suggestions.
Nordita's support during the program on Gravitational Waves from the
Early Universe in Stockholm in 2019 is gratefully acknowledged.
This work was support through grants from the Swedish Research Council
(Vetenskapsradet, 2019-04234).
We acknowledge the allocation of computing resources provided by the
Swedish National Allocations Committee at the Center for Parallel
Computers at the Royal Institute of Technology in Stockholm and
Lindk\"oping.

\end{acknowledgements}

\vspace{2mm}\noindent
{\large\em Software and Data Availability.} The source code used for
the simulations of this study, the {\sc Pencil Code} \citep{2021JOSS....6.2807P},
is freely available on \url{https://github.com/pencil-code/}.
The DOI of the code is https://doi.org/10.5281/zenodo.2315093 {\tt v2018.12.16} 
\citep{axel_brandenburg_2018_2315093}.
The simulation setup and the corresponding data are freely available on
\dataset[doi:10.5281/zenodo.4900075]{https://doi.org/10.5281/zenodo.4900075}; see also
\url{https://www.nordita.org/~brandenb/projects/InflationaryMagnetoGW/}
for easier access of the same material as on the Zenodo site.


\appendix

\section{Relation between $\beta$ and the reheating temperature}
\label{ModelDetails}

At the end of the introduction, we stated that the values of $\beta=7.3$
and $2.7$ are appropriate for a reheating temperatures of $100\GeV$
and $150\MeV$.
Here, we provide the details of this calculation.

To compute $\beta$ for a given reheating energy scale, $T_{\rm r}$,
we follow the formalism of \cite{2017PhRvD..96h3511S}.
The first step is to calculate the Hubble parameter during inflation,
$H_{\rm f}$.
It is related to $T_{\rm r}$ through their Equation~(51), which
we state here in corrected form as
\begin{eqnarray}\label{bound2}
&\ln\left[\displaystyle{\frac{C+D}{\EEEM}
\left(\frac{g_0}{g_r}\right)^{\!\frac{2\alpha}{3}}\!\!
\left(\frac{g_r \pi^2}{30}\right)^{\frac{\!\!4+\alpha}{3}}}\right]\!
+134\alpha\!+(2\alpha+4)\ln\displaystyle{\frac{H_f}{T_r}}\nonumber\\
&-\displaystyle{\frac{7+\alpha}{3}
\ln\left(\frac{3 H_f^2}{8\pi G}\frac{1}{T_r^4}\right)}=0,
\end{eqnarray}
where $C$ and $D$ are functions of $\alpha$ and $\beta$, whose values
are amended by additional $\beta$-dependent factors that take into
account that the spectrum peaks not at the Hubble wavenumber, as assumed
in \cite{2017PhRvD..96h3511S}, but at $k=\kpeak$.
For $\alpha=2$, the numerical values are $C=0.01266\,\beta(\beta+1/2)$
and $D=0.0063\,[\beta(\beta+1/2)/(2\beta+1/2)]^2$.
In \Eq{bound2}, we have corrected the following typo
relative to the expression given by \cite{2017PhRvD..96h3511S}:
the $(C+D)/g_r$ factor in their Equation~(51) is replaced by $(C+D)/(g_r \pi^2/30)$.
By incorporating this correction and combining it with
$(g_r \pi^2/30)^{(7+\alpha)/3}$, their exponent $(7+\alpha)/3$ becomes
$(4+\alpha)/3$ instead; see \Eq{bound2}.
We have also incorporated the parameter $\EEEM$, which represents
the ratio of the electromagnetic energy density to the background.
This value was taken to be unity in Equation~(51) of
\cite{2017PhRvD..96h3511S}.

Given an estimate for $H_f$, the values of $N_r$ and $N$ are obtained
from the expressions
\EQ
N_r=\frac{1}{3} \ln\left[\frac{90\,H^2_{f}}{8 \pi G \pi^2 g_r T_r^4}\right],
\EN
and
\EQ
N+N_r=66.9-\ln\left(\frac{T_r}{H_f}\right)-\frac{1}{3}\ln\frac{g_r}{g_0}.
\EN
which then yields $\beta=\alpha N/N_{\rm r}$; see \Tab{Tbeta2}
for examples considered in this paper.

\begin{table}\caption{
$\beta$ for different values of $T_r$.
}
\begin{center}
\begin{tabular}{cccccc} 
$T_r$&$\EEEM$ & $H_{\rm f}$ [GeV] &$N_r$&$N$&$\beta$\\
\hline
100 GeV&0.01 & $1.56 \times 10^{-8}$&9.3&33.9&7.3\\
100 GeV&0.1 & $5.00 \times 10^{-8}$&10.1&34.4&6.8\\
 150 MeV &0.01 & $1.70\times 10^{-2}$&27.2&36.5&2.7\\
\label{Tbeta2}
\end{tabular}
\end{center}
\end{table}

\section{Initial condition for magnetic and electric energy spectrum during matter-dominated era}
\label{InitCond}

In \Sec{TheModel}, we stated that the initial magnetic energy spectrum
for $k<\kpeak(\eta)$ is proportional to $k^3$.
Here, we provide the details and discuss the $\beta$ dependence of the
steep slope for $k>\kpeak(\eta)$.

We discuss here the initial condition for the numerical simulations in
the post-inflation matter-dominated era.
The initial magnetic and electric field spectra can be understood by
the following argument.
For $\alpha=2$, the magnetic field spectrum is scale-invariant during
inflation.
However, the scale-invariant part does not contribute to the growing
solution in the post-inflation matter-dominated era.
Only to the next order, there is a contribution proportional to $k^3$.
Therefore, for the scales of interest, there is only the $k^3$
spectrum at super-Hubble scales.
Using Equation~(9) and (31) in \citep{2017PhRvD..96h3511S}, it is
concluded that the magnetic energy spectrum for a particular mode $k$
grows with time as $\propto (\eta+1)^{4\beta+2}$ until the mode satisfies
the condition $k (\eta+1) \le \sqrt{2 \beta (2 \beta+1)}$.
Defining $\eta_k$ as the time for which the mode $k$ obeys $k(\eta+1)=\sqrt{2 \beta (2 \beta+1)}$,
the initial condition for the magnetic energy spectrum is then given by
\EQ
    E_{\rm M}\propto \left\{\begin{array}{lr}
    k^3  (\eta+1)^{(4 \beta+2)}, & k \le k_*(\eta), \\
    k^3 (\eta_k+1)^{(4 \beta+2)}\propto k^{1-4 \beta}, & k \ge k_*(\eta).
    \end{array}\right.
\label{EMstep1}
\EN
In the above expression, $\eta_k$ represents the time for the mode $k$
when $k(\eta+1)=1$. Similarly, the initial condition for the electric
energy spectrum is given by
\EQ
    E_{\rm E}\propto \left\{\begin{array}{lr}
    k (\eta+1)^{4\beta}, &
    k \le k_*(\eta), \\
    k (\eta_k+1)^{4\beta}\propto k^{1-4 \beta}, & k \ge k_*(\eta).
    \end{array}\right.
\EN

To get the total magnetic energy density, we can integrate \Eq{EMstep1}
over $k$. Since the magnetic energy spectrum falls very steeply for
$k>\kpeak(\eta)$, taking $\kpeak$ as the upper limit of the integration
is a good approximation.
Thus, we have
\EQ
\int_0^{\kpeak(\eta)}\EM(k) \, \dd k \propto
\kpeak(\eta)^4 (\eta+1)^{4 \beta+2}\propto (\eta+1)^{4 \beta-2},
\EN
which is well obeyed by the numerical results.

\section{Solving the Maxwell equation}
\label{MaxwellEquation}

At the end of \Sec{TheModel}, we described the numerical approach
to solving \Eq{dAk2dt2} and highlighted the analogy to solving
\Eq{d2hdt2}.
Here, we give the details.

In the initial stage of this project, we solved the wave equation for
$\AAA$ with $\sigma=0$ using the default sixth-order finite differences
of the {\sc Pencil Code} using the module {\tt special/disp\_current}.
However, we noticed artificial degrading of the solution at small length
scales---similarly to what was experienced when solving the GW equation
in real space; see \cite{2020GApFD.114..130R} for details.
Therefore, we decided to solve the Maxwell equation in Fourier space.
Assuming $\ii\kk\cdot\tildeAA=\ii\kk\cdot\tildeEE=0$, we have
$\tildeAA''+\sigma\tildeAA'+k^2\tildeAA=0$
with the characteristic equation $\lambda^2+\lambda\sigma+k^2=0$ and
eigenvalues $\lambda_\pm=(-\sigma\pm D)/2$, where $D=\sqrt{\sigma^2-4k^2}$.
If $\sigma=0$, then $D=2\ii k$ and $\lambda_\pm=\pm\ii k$.

\begin{figure}\begin{center}
\includegraphics[width=\columnwidth]{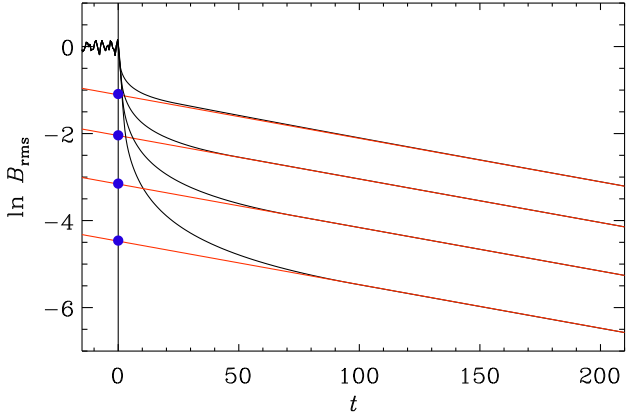}
\end{center}\caption{
Decay of $\ln\Brms$ for a linearly increasing conductivity.
}\label{pcomp_decay}\end{figure}

\begin{figure}\begin{center}
\includegraphics[width=\columnwidth]{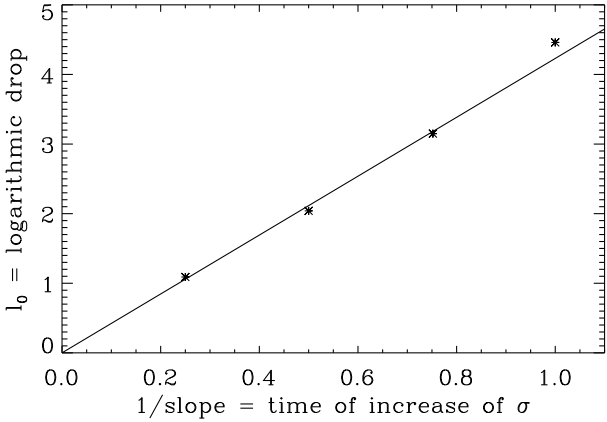}
\end{center}\caption{
Decay of $l=\ln\Brms$ on the slope $s$
for a linearly increasing conductivity.
}\label{plaw}\end{figure}

\begin{figure*}\begin{center}
\includegraphics[width=\textwidth]{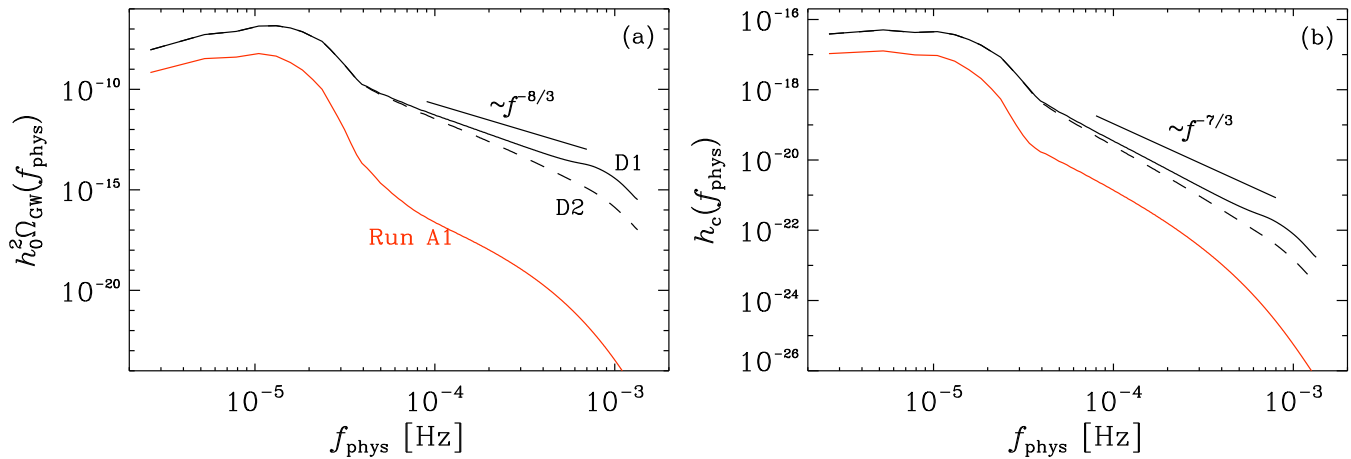}
\end{center}\caption{
(a) $h_0^2\Omega_{\rm GW}(f_{\rm phys})$ and (b) $h_c(f_{\rm phys})$ for Runs~A1 (red lines)
and D1 and D2 (blue lines) for $T_{\rm r}=100\GeV$.
The theoretically anticipated slopes for Kolmogorov-type turbulence
are indicated as well.
}\label{pspecm_strong}\end{figure*}

\begin{table*}\caption{
Comparison between Runs~D1 and D2 with Run~A1.
}\begin{center}
\begin{tabular}{ccc|cc|cc|c|c|cc}
\multicolumn{5}{c}{} &
\multicolumn{2}{c|}{radiation dominated} & $\Brms$ &
\multicolumn{1}{c}{$T_{\rm r}$} &
\multicolumn{2}{c}{scaled to the present time} \\
Run & $\nu$ & $\beta$ & $\EEEM$ & $\EEEl/\EEM$ & $\EEGW$ & $\hrms$ & [$\mu$G] & [GeV] & $\OmGW$ & $h_{\rm c}$ \\
\hline
A1&$1\times10^{-17}$&$6.8$&$0.131$&$8.2$&$4.04\times10^{-4}$&$2.03\times10^{-2}$&$0.48$&$100$&$6.64\times10^{-9}$&$1.62\times10^{-17}$\\
D1&$2\times10^{-21}$&$10$ &$1.4  $&$12  $&$9.91\times10^{-3}$&$8.25\times10^{-2}$&$ 1.3$&100&$1.63\times10^{-7}$&$6.58\times10^{-17}$\\
D2&$5\times10^{-21}$&$10$ &$1.4  $&$12  $&$9.86\times10^{-3}$&$8.27\times10^{-2}$&$ 1.3$&100&$1.62\times10^{-7}$&$6.60\times10^{-17}$\\
\label{Tcomp}\end{tabular}
\end{center}
\end{table*}

Analogously to our solution of the GW equation \citep{2020GApFD.114..130R}
we solve the equation for $\tildeAA$ from one time $t$ to the next $t+\delta t$.
We then make the following ansatz
\begin{eqnarray}
\tildeAA&=&\tildeAA_+ e^{\lambda_+\delta\eta} + \tildeAA_- e^{\lambda_-\delta\eta},\cr
\tildeEE&=&-\tildeAA_+ \lambda_+ e^{\lambda_+\delta\eta} - \tildeAA_- \lambda_- e^{\lambda_-\delta\eta},
\end{eqnarray}
where the coefficients $\tildeAA_+$ and $\tildeAA_-$ are determined from
$\tildeAA$ and $\tildeEE$ at the time $\eta$.
We can then write the solution for the time $\eta+\delta\eta$ in matrix form as
\begin{equation}
\pmatrix{
{\tildeAA}\cr\tildeEE}_{\eta+\delta\eta}=
\pmatrix{c_A & s_A \cr s_E & c_E }
\pmatrix{\tildeAA\cr\tildeEE}_\eta,
\end{equation}
where
\begin{equation}
\MMMM\equiv \pmatrix{c_A & s_A \cr s_E & c_E }
=\pmatrix{
\cos k\delta\eta& -k^{-1}\sin k\delta\eta\cr
k \sin k\delta\eta& \cos k\delta\eta}
\end{equation}
is a rotation matrix for $\sigma=0$, and
\begin{equation}
\MMMM={1\over D}\pmatrix{ 
\lambda_+ e^{\lambda_-\delta\eta} - \lambda_- e^{\lambda_+\delta\eta} &
          e^{\lambda_-\delta\eta} -           e^{\lambda_+\delta\eta} \cr
\lambda_+ \lambda_- (e^{\lambda_+\delta\eta} - e^{\lambda_-\delta\eta} ) &
\lambda_+ e^{\lambda_+\delta\eta} - \lambda_- e^{\lambda_-\delta\eta} }
\end{equation}
in the general case $\sigma\neq0$.
This solution is now implemented in the new module {\tt magnetic/maxwell}.

\section{Conductivity changes}
\label{ConductivityChanges}

At the end of \Sec{EvolStepII}, we emphasized that both for $\sigma=0$
and $\sigma\to\infty$, magnetic fields are undamped, but that there can
be strong decay for intermediate values.
Here, we demonstrate that this decay depends on the duration $T$
of the transition from $\sigma=0$ to $\sigma\to\infty$.

When $\sigma=0$, there are electromagnetic waves that remain undamped.
For $\sigma\neq0$, however, the magnetic field experiences
magnetic diffusion such that $\Brms\propto\exp(-t k^2/\sigma)$,
where $t$ is ordinary time.
Only for $\sigma\to\infty$ can the magnetic field survive.
Thus, we expect that a slow transition from $\sigma=0$ to
$\sigma\to\infty$ can result in a significant loss of
magnetic energy.

To quantify the transitional loss of $\Brms$ when $\sigma$ is not yet
large enough, we assume a linear conductivity increase of the form
$\sigma=t/T$ with a transition time $T$ after which $\sigma$ reaches a
value $\sigma_{\max}$ such that $\Brms$ follows a slow exponential decay
at late times.
We determine the logarithmic drop, $\Delta\ln\Brms$ and
extrapolate its value back to the time $t_0$ when
$\sigma$ was still constant; see \Fig{pcomp_decay}.
\Fig{plaw} shows the dependence $\Delta\ln\Brms$ versus $T$,
and is seen to follow an approximately linear behavior.
For small $T$, the losses are small.
This justifies our approach of assuming an instantaneous switch
from $\sigma=0$ to $\sigma\to\infty$.

\section{Stronger field strengths}
\label{StrongerField}

In \Sec{EvolStepII}, we discussed the possibility of a Kolmogorov-type
scaling for large magnetic energies, which could imply an
$f_{\rm phys}^{-8/3}$ scaling for $h_0^2\Omega_{\rm GW}(f_{\rm phys})$
and an $f_{\rm phys}^{-7/3}$ scaling for $h_c(f_{\rm phys})$; see
\cite{2020PhRvD.102h3512R}.
\Fig{pspecm_strong} shows a comparison between Run~A1 and two new ones,
Runs~D1 and D2, for which the initial electromagnetic energies are
unphysically large and even exceeding unity; see \Tab{Tcomp}.
Runs~D1 and D2 differ in the values of the viscosity ($2\times10^{-4}$
and $5\times10^{-4}$, respectively), but in both cases they are still
larger than the value used in Run~A1 ($10^{-4}$).
This is because of the stronger magnetic field strength in these
runs, which requires larger viscosity and magnetic diffusivity.

\section{Avoiding the discontinuity at the end of reheating}
\label{Avoiding}

\begin{figure}\begin{center}
\includegraphics[width=\columnwidth]{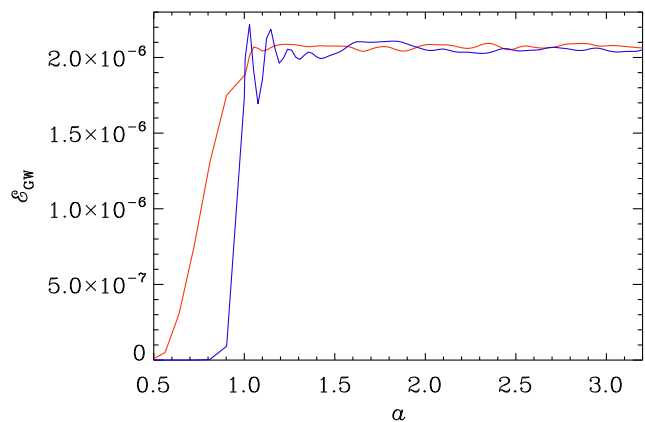}
\end{center}\caption{
Comparison of $\EEGW(t)$ for a run with quenching (red line)
and without (blue).
}\label{pcomp_befaft2_EEGW_quench}\end{figure}

In \Sec{DifferentBeta}, we discussed whether the discontinuities
in $a''/a$ and $f''/f$ could be responsible for the occurrence of
oscillations.
In \Fig{pcomp_befaft2_EEGW_quench}, we compare two runs, where in one
case the discontinuities have been smoothed out.
The original run corresponds here to Run~D2 of \cite{Bran+He+Shar21},
which differs from the present runs in that this one has helicity.
This also leads to a much larger value of $\kpeak(1)$ of about 18
for $\beta=7.3$.
The smoothing of the discontinuities has been accomplished by dividing
the two ratios by a quenching term $Q=1+(a/a_{\rm c})^n$, where $n=20$
has been chosen to make the onset of quenching sharp, and $a_{\rm c}$
has been chosen so as to ensure that $1/Q$ is small enough well before
$\eta=1$ has been reached, i.e., when $a\approx a_{\rm c}$.
Generally, the quenching leads to a decrease of GW energy by the end
of the run.
We have therefore rescaled the spectra so as to have comparable
spectral amplitudes.
Looking at \Fig{pcomp_befaft2_EEGW_quench}, we see that the growth
of $\EEGW$ slows down before $\eta=1$ is reached and that the
oscillations are now absent (red line).

\begin{figure}\begin{center}
\includegraphics[width=\columnwidth]{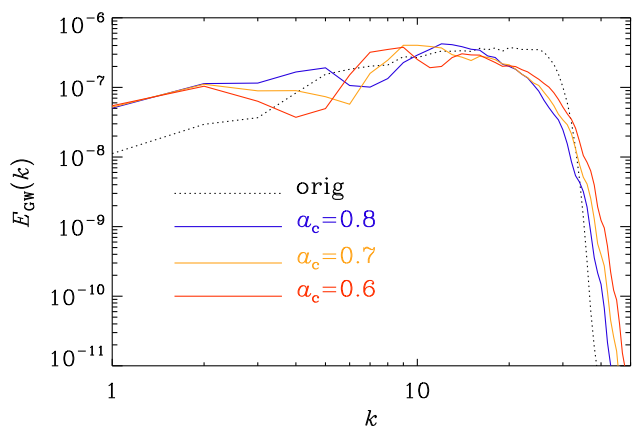}
\end{center}\caption{
GW energy spectra for different values of $a_{\rm c}$, compared with
the original Run~D2 of \cite{Bran+He+Shar21}.
}\label{pspec34}\end{figure}

In \Fig{pspec34}, we show GW spectra for different values of $a_{\rm c}$.
We see that the effect of quenching is to make the spectra somewhat shallower.
This suggests that a correspondence between the stress
spectra and the spectra of GW energy can only be reached when the
rapid growth of GW energy has come to an end.
Denoting the computation of spectra again by an operator $\Sp$,
we can say that $\Sp(T)\approx\Sp(k^2\tilde{h})=k^2\Sp(k\tilde{h})=
\EGW(k)$, which agrees with our findings.
Thus, in conclusion, the change from a $k^1$ spectrum to $k^0$ found in
the simulations of \cite{Bran+He+Shar21} occurs when the growth of
electromagnetic energy has stopped.
This is when $f'=f''=0$, but it is not a direct consequence of the
discontinuity at $\eta=1$ and therefore not an artifact.

\bibliography{ref}{}
\bibliographystyle{aasjournal}
\end{document}